\documentclass[prd,superscriptaddress,onecolumn,showpacs,11pt,msmath,
preprintnumbers,showkeys]{revtex4}
\usepackage{wrapfig,rotating}
\usepackage{graphicx,epsfig,color}



\usepackage{graphicx}
\usepackage{dcolumn}
\usepackage{bm}

\def\beq{\begin{equation}}
\def\eeq{\end{equation}}
\def\beeq{\begin{eqnarray}}
\def\eeeq{\end{eqnarray}}

\newcommand \pt {$p_T$}
\def\effs {$\sigma_{eff}$}
\def\2GPD{$_2\mbox{GPD}$}

\def\12{$1\otimes 2$}
\def\22{$2 \otimes 2$}

\def\Qsep{Q_{\mbox{\rm\scriptsize sep}}}
\def\Qsep2{Q^2_{\mbox{\rm\scriptsize sep}}}
\newcommand*{\relpt}{$\Delta^{rel}_{soft}p_T$}

\begin{document}

\title{Dynamical approach to MPI four-jet production in Pythia}

\author{B.\ Blok$^{1}$,
 P.\ Gunnellini$^{2}$
\\[2mm] \normalsize $^1$ Department of Physics, Technion -- Israel Institute of Technology,
Haifa, Israel\\
\normalsize $^2$ Deutsches Elektronen-Synchrotron (DESY), Notkestra$\ss$e 85, 22761 Hamburg, Germany}

\begin{abstract}
 We 
  modify
  the treatment of Multiple Parton Interactions (MPI) in \textsc{Pythia} by including the \12 mechanism and treating the \22 mechanism in a model-independent way. The \22 mechanism is calculated within the mean field approximation, and its parameters are
  expressed through Generalized Parton Distributions extracted from HERA data. The parameters related to the transverse parton distribution inside the proton are thus independent of the performed fit. The \12 mechanism is included along the lines of the recently developed formalism in perturbative QCD. A unified description of
  MPI at moderate and hard transverse momenta is obtained within a consistent framework, in good agreement with experimental data measured at 7 TeV. Predictions are also shown for the considered observables at 14 TeV. The corresponding code implementing the new MPI approach is made available.
 \end{abstract}

  \maketitle
\thispagestyle{empty}

\vfill
\section{\bf Introduction}

It is widely realized now that hard {\em Multiple Parton Interactions}\/ (MPI)  play an important role in the description of inelastic proton-proton ($pp$) collisions at high center-of-mass energies.
Starting from the eighties~\cite{TreleaniPaver82,TreleaniPaver85,mufti,dDGLAP} until the last decade~\cite{Treleani,Diehl,DiehlSchafer,Diehl2,Wiedemann,Frankfurt,Frankfurt1,SST,stirling,stirling1,Ryskin,Berger,BDFS1,BDFS2,BDFS3,BDFS4,Gauntnew,Gauntadd,gieseke1,gieseke2,gieseke3,Sjodmok}, extensive theoretical studies have been carried out. Attempts have been made to incorporate multi-parton collisions in Monte Carlo (MC) event generators \cite{Sjostrand:2007gs,Corke:2011yy,Corke:2010yf,Herwig,Lund}. Multiple parton interactions can serve as a probe for {\em non-perturbative correlations}\/ between partons in the nucleon wave function and are crucial for determining the structure of the Underlying Event (UE) at Large Hadron Collider (LHC) energies. Moreover, they constitute an important background for new physics searches at the LHC. A large number of experimental measurements have been performed at the Tevatron \cite{Tevatron1,Tevatron2,Tevatron3} and at the LHC \cite{Atlas,cms1,cms2,Chatrchyan:2013qza}, showing evidence for MPI at both soft and hard scales. This latter case is usually referred to as ``Double Parton Scattering'' (DPS), which involves two hard scatterings within the same hadronic collision. The cross section of such an event is generally expressed in terms of the $\sigma_{\rm eff}$.
In the mean field approximation \effs~\cite{TreleaniPaver82,TreleaniPaver85,mufti,dDGLAP,Treleani,Diehl,DiehlSchafer,Diehl2,Wiedemann,Frankfurt,Frankfurt1,SST,stirling,stirling1,Ryskin,Berger,BDFS1,BDFS2,BDFS3,BDFS4,Gauntnew,Gauntadd,Sjodmok}, is the effective area which measures the transverse distribution of partons inside the colliding hadrons and their overlap in a collision.

\par Recently, a new approach based in perturbative Quantum Chromodynamics (pQCD) has been developed \cite{BDFS1,BDFS2,BDFS3,BDFS4} for describing the MPI and its main ingredients are:
\begin{itemize}
\item the MPI cross sections are expressed through new objects, namely double Generalized Parton distributions (GPD$_2$);
 \item besides  the conventional mean field parton model approach to MPI, represented by the so-called \22 mechanism (see Fig. 1 left),
  an additional \12 mechanism (Fig. 1 right) is included.  In this mechanism, which can be described in pQCD, the parton from one of the nucleons splits at some hard scale and creates two hard partons that may participate in MPI. This mechanism leads to a significant transverse-scale dependence of MPI cross sections;
 \item the contribution of the \22 mechanism to  GPD$_2$  is calculated in a mean field approximation with model-independent parameters.
\end{itemize}
\par The use of this new formalism at LHC experiments needs its implementation in MC event generators which has not been performed yet. The purpose of the present paper is to make a step ahead towards the implementation of this formalism into MC generators. We use the standard simulation of the MPI implemented in \textsc{Pythia}~\cite{Corke:2010yf}, but with values of \effs\ calculated by using the QCD-based approach of \cite{BDFS1,BDFS2,BDFS3,BDFS4}, i.e. including \12 processes.

\begin{figure}[htbp]
\begin{center}
\includegraphics[scale=0.65]{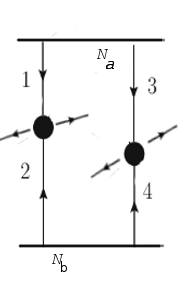}\hspace{3cm}
\includegraphics[scale=0.65]{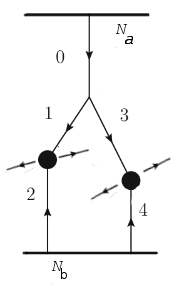}\\
\caption{Sketch of the two considered MPI mechanisms: \22 (left) and \12 (right) mechanism.}
\end{center}
\end{figure}

\par The current approach used for the description of the MPI in \textsc{Pythia} is based on \cite{Corke:2011yy,Corke:2010yf}.
The \textsc{Pythia} code uses parton distribution functions, dependent on the impact parameter of the collision.
From the theoretical point of view these are just one-particle Generalized Parton Distributions GPD$_1$ (see e.g. \cite{DiehlS,radyushkin} for a review).
The parameters set in the \textsc{Pythia} simulation relative to the transverse parton density are extracted from fits to experimental data on UE, sensitive to the contribution of the MPI. This procedure is closely related to mean field based schemes, see e.g.~\cite{BDFS1}.
\par Such an approach has, however, a number of difficulties, both conceptual and practical. First of all, a problem arises at the level of mean field approximation. The transverse parton distributions have been extracted from $J/\Psi$ photoproduction measurements  at the HERA collider, using QCD factorisation theorems \cite{Frankfurt,Frankfurt1,DiehlS,radyushkin}. Hence they can not be treated as free parameters of the model. Secondly, it has been observed that different \textsc{Pythia} parameters are obtained when data sensitive to a different region of the MPI spectrum are used for the fits. For example, it has been shown \cite{gunnelini} that different parameters result for fits to UE or hard MPI data. This might be an indication that an additional transverse scale dependence, which is not present in the mean field approach, is needed to describe experimental data on UE and hard MPI simultaneously. Recent improvements in the \textsc{Pythia} MPI model include a dependence of the parton transverse density on the longitudinal momentum fraction ($x$)~\cite{Corke:2011yy}, but this only accounts for the $x$ values of the hardest dijet. A complete $x$ dependence which considers soft and hard partons may be irrelevant for the UE description where the transverse scales are rather close, but may become important for measurements sensitive to hard MPI.
\par The approach used in this paper combines the standard \textsc{Pythia} MPI model with the one of \cite{BDFS1,BDFS2,BDFS3,BDFS4}. We use a single gaussian to model the matter distribution function of the protons in \textsc{Pythia}. With these settings, the value of $\sigma^{(0)}_{eff}$ would be constant and independent on the scale. In order to implement the $x$ and the scale dependence of \effs\ in collisions where a hard MPI occur, these events are rescaled according to:
\beq
\sigma_{eff}=\frac{\sigma^{(0)}_{eff}}{1+R},
\eeq
where $\sigma^{(0)}_{eff}$ is the effective cross section in the mean field approach calculated in a model independent way from GPD$_1$, parameterized from HERA data \cite{Frankfurt,Frankfurt1,BDFS1}, and $R$ corresponds to the correction due to \12 mechanism \cite{BDFS3,BDFS4}.
Such an approach is equivalent to using the GPD$_1$-based transverse parton densities for parton transverse distributions.\\
\par The main result of this paper is that the approach discussed above gives a unified description of both hard MPI and UE experimental data, with good accuracy and few fit parameters. The fit parameters are related to the amount of simulated MPI and of the color string reconnection, and to the separation scale between soft- and hard-scale processes, $Q_0^2$ whose value is expected to lie in the range 0.5-2 GeV$^2$. The transverse scale dependent function $R$ is calculated numerically by solving the nonlinear evolution equation \cite{BDFS3,BDFS4}. Predictions using this approach are shown later in the paper, and are labeled as ``UE Tune Dynamic \effs''. Our analysis shows that the values of observable for UE are quite close to the results obtained in a free parton model (mean field approximation), while
the inclusion of transverse scale dependent rescaling calculated in pQCD \cite{BDFS3} improves the description of hard MPI.

\par The paper is organized as follows. In section 2, basic theoretical ideas of the used approach are presented, while in section 3 their MC implementation is discussed. In section 4, comparisons for various predictions to observables measured at 7 TeV are shown. In section 5, predictions for these observables are presented for $pp$ collisions at 14 TeV. In section 6
we compare our approach with the recently developed  HERWIG-EE-5 approach, before drawing the conclusions in Section 7.

\section{ A summary of the theoretical background}
 The MPI four-jet cross section is characterized by the cross section \effs, which corresponds to an effective interaction area \cite{BDFS1}, and can be written as:
\beq
\frac{d\sigma^{\textrm{\tiny four-jet}}}{dt_{12}dt_{34}} = \frac{d\sigma^{\textrm{\tiny two-jet}}}{dt_{12}} \frac{d\sigma^{\textrm{\tiny two-jet}}}{dt_{34}}\times \frac{1}{\sigma_{eff}},
\eeq
where partons 1 and 2 create the first (12), and partons 3 and 4 the second (34) dijet. The pQCD calculation leads to the following expression for \effs\ in terms of two-particle GPD:
\begin{eqnarray}
\frac1{\sigma_{eff}} &\equiv&
\int \frac{d^2\vec{\Delta}}{(2\pi)^2}[\mbox{ } _{[2]}G_2(x_1,x_3, Q_1^2,Q_2^2;\vec\Delta) _{[2]}G_2(x_2,x_4, Q_1^2,Q_2^2; -\vec\Delta)\nonumber\\[10pt]
&+&_{[1]}G_2 (x_1,x_3, Q_1^2,Q_2^2;\vec\Delta)
{}_{[2]}G(x_2,x_4, Q_1^2,Q_2^2;-\vec\Delta)\nonumber\\[10pt]
 &+& {}_{[1]}G_2(x_2,x_4, Q_1^2,Q_2^2;\vec\Delta){}_{[2]}G_2(x_1,x_3, Q_1^2,Q_2^2;-\vec\Delta)].\label{2}
 \end{eqnarray}
The second and third terms in Eq.~\ref{2} correspond to the \12 mechanism, when two partons are generated from the splitting of a parton at a hard scale after evolution, while the first term corresponds to the conventional case of two partons evolving from a low scale, namely the \22 mechanism. This first term can be calculated in the mean field approximation \cite{Frankfurt,Frankfurt1,BDFS1}. The momentum $\Delta$ is conjugated to the relative distance between the two participating partons. The full double GPD is a sum of two terms:
\beq
G_2(x_1,x_3,Q_1^2,Q_2^2,\Delta)=_{[1]}G_2(x_1,x_3,Q_1^2,Q_2^2,\Delta)+_{[2]}G_2(x_1,x_3,Q_1^2,Q_2^2,\Delta)\label{3}.
\eeq

\par Here $_{[2]}G_2$ corresponds to the part of double GPD$_2$, when both partons are evolved from the initial nonperturbative scale, while $_{[1]}G_2$ corresponds to the case when one parton evolves up to some hard scale, where it splits to two successive hard partons, each of them in turn participating to the hard dijet event.
We refer the reader to \cite{BDFS1,BDFS2} for the detailed definitions of $_{[1]}G_2$ and $_{[2]}G_2$ and their connection to light cone wave functions of the nucleon.\\

For the two-parton GPD$_2$ we have:
\beq
_{[2]}G_2(x_1,x_3,Q_1^2,Q_2^2,\Delta)=D(x_1,Q_1)D(x_3,Q_2)F_{2g}(\Delta,x_1)F_{2g}(\Delta,x_3),
\eeq

where $D(x,Q^2)$ is a conventional parton distribution function (PDF). The use of the mean field approximation results in:
\beq
_{[2]}G_2(x_1,x_3,Q_1^2,Q_2^2,\Delta)=G_1(x_1,Q_1^2,\Delta)G_1(x_1,Q_1^2,\Delta),\label{slon}
\eeq
and
\beq
G_1(x_1,Q_1^2,\Delta)=D(x_1,Q_1)F_{2g}(\Delta,x_1).\label{slon1}
\eeq
For the two gluon form factor $F_{2g}$, we use the exponential parametrization \cite{Frankfurt1}. In fact, it leads to the same numerical results as the dipole form \cite{Frankfurt}, but it is more convenient for calculations. This parametrization is unambiguously fixed by $J/\Psi$ diffractive charmonium photo/electro production at HERA. The functions $D$ are the conventional nucleon structure functions and $F_{2g}$ can be parameterized as:
\beq
F_{2g}(\Delta,x)=\exp(-B_g(x)\Delta^2/2),
\label{d1}
\eeq
where $B_g$($x$)= $B_0$ + 2$K_Q\cdot\log(x_{0}/x)$, with $x_0\sim 0.0012$, $B_0=4.1$ GeV$^{-2}$ and $K_Q=0.14$ GeV$^{-2}$. In our implementation the central values of the parameters $B_0$ and $K_Q$ \cite{Frankfurt1} have been used, which are known with an accuracy of $\sim 8\%$. Integrating over $\Delta^2$, we obtain for the part of \effs\ corresponding to the first term in Eq.~\ref{2}:
\beq
\frac{1}{\sigma^{(0)}_{eff}}=\frac{1}{2\pi}\frac{1}{B_g(x_1)+B_g(x_2)+B_g(x_3)+B_g(x_4)}.\label{mura}
\eeq
where $x_{1..4}$ are the longitudinal momentum fractions of the partons participating in the \22 mechanism. This cross section corresponds to the free parton model and is model independent in the sense that its parameters are determined not from the fit of experimental LHC data, but from the fit of single parton GPD$_1$.  The maximum transversality kinematics i.e. $4Q^2=x_1x_2s$ for each dijet, have been considered in our approach, being $Q$ the dijet transverse scale, and $x_1,x_2$ the Bjorken fractions of the jets.
\par The second and third terms in Eq. \ref{2} are parameterized as:
\beq
\sigma_{eff}=\frac{\sigma^{(0)}_{eff}}{1+R}\label{scaledep},
\label{mur}
\eeq
where $R(Q_1^2,Q_2^2,Q_0^2)$  is calculated by solving iteratively the nonlinear evolution equation, as explained in detail in \cite{BDFS3,BDFS4}. According to the results of \cite{BDFS4}, the dependence of $R$ on $x_i$ in the maximum transversality regime is very weak and can be neglected with high accuracy. The function $R$ also depends on the physical parameter $Q_0^2$ which corresponds to the separation scale between soft and hard dynamics where the GPD$_2$ is assumed to factorize.

\section{Monte Carlo implementation and definition of experimental observables}
\par In this paper we carry out two types of simulations: one based on the new approach defined in the sections 1 and 2 and one which follows the standard \textsc{Pythia} approach, used for comparison.\\

\par Let us recall the standard \textsc{Pythia} approach which is referred as to "UE tune" hereafter. In this study we use the \textsc{Pythia}~8.185 Monte Carlo event generator \cite{Sjostrand:2007gs}. It simulates a 2$\rightarrow$2 matrix element interfaced to parton shower and Underlying Event (UE).  The \textsc{Pythia}~8 event generator uses a simulation of the parton shower ordered in transverse momentum and  the Lund string model \cite{Andersson:1998tv} to implement the hadronization process. The performed study has considered as a starting point the UE simulation implemented in the \textsc{Pythia}~8 tune 4C \cite{Corke:2010yf}. This simulation makes use of the CTEQ6L1~\cite{Pumplin:2002vw} parton distribution function and of a simple gaussian as a transverse matter distribution function. A fit to experimental data sensitive to the UE is performed in order to optimize the parameters related to the amount of MPI and colour reconnection in the simulation. The fit operation has been carried out by using the \textsc{RIVET} \cite{Buckley:2010ar} software, combined with the \textsc{PROFESSOR} machinery \cite{Buckley:2009bj}. For the tune, two different observables have been considered at a center-of-mass energy of 7 TeV measured by the ATLAS experiment \cite{Aad:2010fh}. They are related to the multiplicity, N$_{chg}$, and the sum of the transverse momentum, $\sum {\rm p}_{rm T} $, of the charged particles in the region transverse to the direction of the leading charged particle in each event. The performed fit has used only the data points corresponding to transverse momenta of the leading charged particle between 2.0 and 15.0 GeV. The exclusion of the very low $p_T$ region~($\le 2$ GeV) is motivated by the fact that processes at those scales are expected to be dominated by soft physics, including diffractive processes and soft nonperturbative correlations, i.e. along the lines of \cite{BDFS2}.
The upper cut off is arbitrary, since its variation starting from 5 GeV does not change the values of the observables.

The result of the fit consists of a new set of UE parameters implemented in the ``UE tune'' hereafter. The values of the \textsc{Pythia}~8 parameters obtained for the ``UE tune'' after the fit are shown in Table \ref{table1}.

\begin{table}[htbp]
\begin{tabular}{c c} \hline
 \textsc{Pythia}~8 Parameter & Value obtained for the UE tune\\\hline
 MultipleInteractions:$p_T^0$Ref & 2.659\\
 BeamRemnants:reconnectRange & 3.540\\\hline\hline
 {Reduced $\chi^2$} & 0.647\\ \hline
 {\effs\ (7 TeV)} (mb) & 29.719\\
 {\effs\ (14 TeV)} (mb) & 32.235\\ \hline
\end{tabular}
\caption{\textsc{Pythia}~8 parameters obtained after the fit to the UE observables. The value of pT0Ref is given at a reference energy of 7 TeV. The values of the reduced $\chi^2$ and of \effs\ at 7 and 14 TeV are also shown in the table. }
\label{table1}
\end{table}

The first
parameter listed in the table
 refers to the value of transverse momentum, $p_T^0$, defined at $\sqrt{s}=7$ TeV, used for the regularization of the cross section in the infrared limit, according to the formula $1/p_T^4\rightarrow 1/(p_T^2+p_{T}^{0\mbox{ }2})^2$. The second parameter is the probability of color reconnection among parton strings. The value of \effs\ is found to be around 29.7 mb at 7 TeV; this value is significantly smaller than the one obtained by tuning the correlation observables of the four-jet scenario \cite{Chatrchyan:2013qza}, which is around 19-21 mb. Note that the value of 29.7 mb is quite close to the one determined in
 mean field approach \cite{BDFS1,BDFS4}.

After fitting the UE observables for the ``UE tune'' determination, the considered predictions are also tested against measurements sensitive to the hard spectrum of the MPI. Measurements of such type have been conducted by studying correlations between outgoing objects in a proton-proton collision, for instance in four-jet final states measured at 7 TeV by CMS \cite{Chatrchyan:2013qza}. In this scenario, two dijets have been selected at different transverse momentum; two jets are required to have \pt\ larger than 50 GeV and they are classified as ``hard-jet pair'', while the so-called ``soft-jet pair'' is composed by the two other jets selected with \pt\ greater than 20 GeV. Two correlation observables, $\Delta S$ and \relpt\, that are sensitive to DPS, have been considered. They are, respectively, the azimuthal angle between the two dijet planes and the \pt\ balance between the soft jets and are defined as follows:
\begin{equation}\label{dels}
\Delta S=\arccos\left(\frac{\vec{p}_T(pair_1)\cdot \vec{p}_T(pair_2)}{|\vec{p}_T(pair_1)|\times |\vec{p}_T(pair_2)|}\right),\\
\end{equation}
\begin{equation}\label{deltarel}
\Delta^{rel}p_T = \frac{|\vec{p}_{T}^{\textrm{ }jet_1}+\vec{p}_{T}^{\textrm{ }jet_2}|}{|\vec{p}_{T}^{\textrm{ }jet_1}|+|\vec{p}_{T}^{\textrm{ }jet_2}|},\\
\end{equation}
where $pair_1$ ($pair_2$) is the hard (soft) jet pair and $jet_1$ ($jet_2$) is the leading (subleading) soft jet.\\
\par Let us now move to the new approach, based on the dynamical pQCD-based formalism, described in sections 1 and 2.
The $x$ and scale dependence of \effs\ has been implemented by reweighting on an event-by-event basis the Monte Carlo simulation in presence of a hard and moderate MPI. The $x$ dependence is given by Equation~\ref{mura}, where $x_{1,2} $ are taken as the longitudinal momentum fractions of the partons participating in the hardest scattering, while $x_{3,4} $ refer to the longitudinal momentum fractions of the partons participating in the hardest MPI. The scale dependence is expressed by Equation~\ref{scaledep}, where $R$ takes for $Q_1$ and $Q_2$ the scales of, respectively, the hard scattering and of the hardest MPI. Different values of $Q_0^2$ have been considered in the range between 0.5 and 2 GeV$^2$.
\par We considered both the case of moderate MPI (i.e. MPI at scales of several GeV), relevant for UE, and hard MPI.
\par  For UE we treat separately the events where there is only one hard scattering, which  are not rescaled, and the events with additional hard MPI. For the latter events two approaches were checked. First, we rescaled these events according to Equations \ref{mura} and \ref{scaledep}, taking as $Q_1$ and $Q_2$ for the $R$ function the scales of the two hardest scatterings. As shown in Section IV, the influence of this rescaling is very small (less than 5$\%$), with respect to the standard \textsc{Pythia} "UE tune". This may be connected both with the small values of $R$ obtained for UE, and with the fact that the ladder splitting is roughly taken into account for such scales by the large value of the parameter $p^0_{T}\sim 2$ GeV. \\Subsequently, the second approach tried was to rescale only MPI events starting from the scale of order 4-5 GeV. When we rescale only the MPI starting from this (or a higher) scale, UE observables are not affected at all. At the same time, with this approach we avoid possible double counting effects, since at these scales the regularization formula in \textsc{Pythia} represents an ansatz for higher twist effects, including MPI. Thus, while using \textsc{Pythia}, we can neglect rescaling of MPI in UE, fitting $p^{0}_{T}$ instead. With the current accuracy, any of these two approaches can be used, leading to identical numerical results.
This is in agreement with the approach documented in \cite{BDFS4}, where it was argued that at scales relative to UE the values of \effs\ are close to the ones calculated in the mean field approximation.
\par We consider now the case of hard MPI, specifically DPS. Two different processes may produce four jets in the final states. The first one is the so-called Single Parton Scattering (SPS) where the four jets are emitted through the same chain while the second one is DPS where the two hard interactions produce one dijet each. A different  event topology is expected from these processes: if the four jets are produced through SPS, a high correlation between the objects of the final state is present and this is reflected in their relative configuration in the transverse plane. The direction of the hard jets, for example, is randomized by the emission of the additional two jets within the same chain and their initial $p_T$ balance is ruined. Instead, jet pairs coming from DPS events, namely from two independent scatterings, tend to be uncorrelated and their initial back-to-back configuration is less subject to smearing effects coming from additional hard radiation: the jet pairs are expected to exhibit a more balanced configuration in $p_T$ and azimuthal angle. In particular, as shown in \cite{Chatrchyan:2013qza}, DPS events add a relevant contribution at low values of $\Delta$S and \relpt. Here we consider the experimentally relevant example, when the two dijet scales are 50 and 20 GeV.

Similarly as before, the $x$ and scale dependence of \effs\ have been implemented by reweighting on an event-by-event basis the Monte Carlo simulation, as explained above.

 In case that only MPI with $p_T$ scales smaller than 15 GeV are present in the collision, no $x$ and scale dependence is applied to the \effs\ value of the corresponding event. The choice of 15~GeV is motivated by the fact that we need to treat differently the two contributing processes, SPS and DPS. Events where the two dijets are produced through SPS accompanied by moderate MPI, should not be reweighted~\cite{BDFS1,stirling1,DiehlSchafer,BDFS2,BDFS3}. In case a hard MPI occurs in the collision, dynamical \effs\ values are used. In this way, we assume that all collisions with a MPI scale greater than 15 GeV produce the second hard dijet $p_T>20$ GeV pair selected in the considered four-jet scenario, while MPI at lower scales are below threshold for producing jets with $p_T$ $>$ 20 GeV. This approach is generally followed by standard experimental measurements for \effs\ determination, as the ones documented in \cite{Atlas,cms1}. For our studies, lowering the 15 GeV cut off by 5-10 GeV shows variations of the predictions of DPS-sensitive observables of less than $2\%$. This is a clear indication of the consistency of our approach. \\

Various simulation settings have been considered for comparison:

\begin{itemize}
\item ``UE tune'': predictions obtained with the parameters listed in Table \ref{table1} and without applying any reweighting of the simulation; this tune uses a constant value of \effs, following the standard \textsc{Pythia} approach;
\item ``UE tune Q$^2$-dep'': predictions obtained with the UE parameters listed in Table \ref{table1} and by applying the scale dependence of \effs\ with $Q_0^2$ = 1 GeV$^2$;
\item ``UE tune $x$-dep'': predictions obtained with the UE parameters listed in Table \ref{table1} and by applying the $x$ dependence of \effs;
\item ``UE tune Dynamic \effs'': predictions obtained with the UE parameters listed in Table \ref{table1} and by applying both the $x$ and the scale dependence with $Q_0^2$ = 1 GeV$^2$.
\end{itemize}

For the considered ``UE tune Dynamic \effs'', predictions using $Q_0^2$ values equal to 0.5, 1 and 2 GeV$^2$ have been also tested and compared.

\par A full MC implementation of the presented approach may be different from the one used in this paper, which relies on reweighted events simulated by \textsc{Pythia}. There are at least three  reasons for it:
\begin{itemize}
\item by using the \textsc{Pythia} event generator, all ladders are assumed to evolve independently from the low transverse scale, the initial-state radiation (ISR) being regularized by primordial gluon distribution with a transverse scale equal to $p_T^0$. No parton ladder splittings are included in this approach;
\item in \textsc{Pythia}, the geometric picture of the collisions in the impact parameter space corresponds to the \22 mechanism, while for \12 mechanism the geometrical picture would be different;
\item for multi MPI events, namely for events with several MPI within the same collision,  we neglect the change of relative weight of \12 and \22 mechanisms.
\end{itemize}

In this paper, these effects are neglected. First, the good agreement with experimental data shows that the high regularization scale $p_T^0$ may be a good alternative parametrization of the ladder splittings and of the corresponding changes in ISR at the UE transverse scales. In other words, for UE the high $p^0_{T}$, which regularizes the charged particle multiplicity, also approximately fits the change of multiplicity due to ladder splitting. The \textsc{Pythia} regularization formula in this case can be viewed as an ansatz for twist expansion, that may include part of the MPI.  Note that the ladder splitting scale is much smaller than the scales of hard dijets created by partons that evolve after the splitting \cite{BDFS4,Gauntnew}. So the effective ladder splitting is partly taken into account for UE by a high $p^0_{T}$ value. This is the reason why the UE observables change only slightly in the new approach. On the other hand for hard MPI, when the hard splitting scale is much larger than $p^0_{T}$, the inaccuracy in accounting for ISR at small $p_T$ can be safely neglected.
\par Second, the direct calculation along the lines of \cite{SST} shows that neglecting the change of geometrical picture and of the relative weight between mean field and \12 mechanisms, when more than two separate dijet events are present, does not lead to numerical changes.

We conclude that using events simulated with \textsc{Pythia} and reweighted with $x$- and scale-dependent values of \effs\ is a good approximation. In this way, we investigate the influence of changes of \effs\ on MC observables sensitive to UE and DPS.\\

\section{Results for  7 TeV}
\begin{figure}[htbp]
\begin{center}
\includegraphics[scale=0.65]{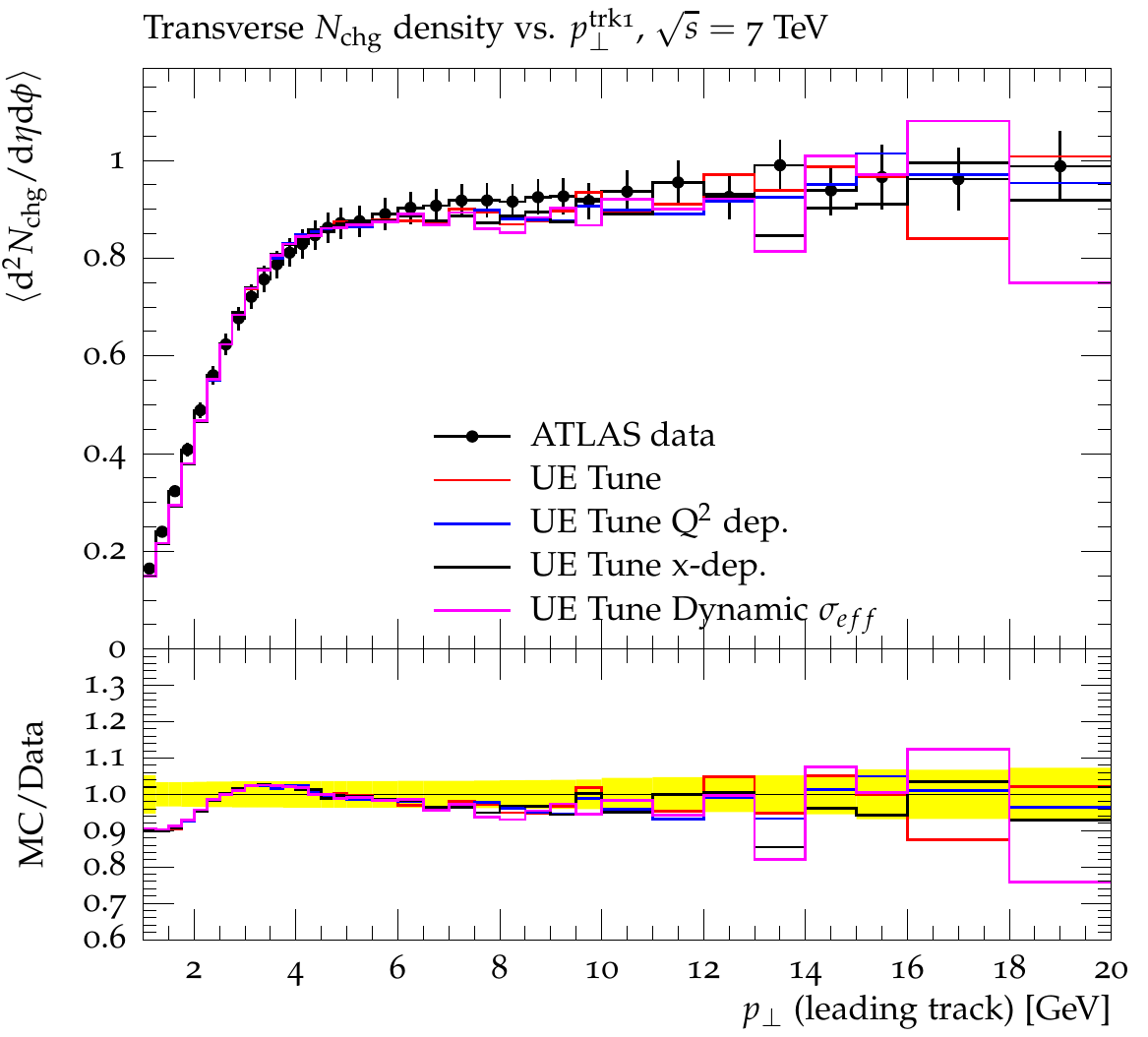}
\includegraphics[scale=0.65]{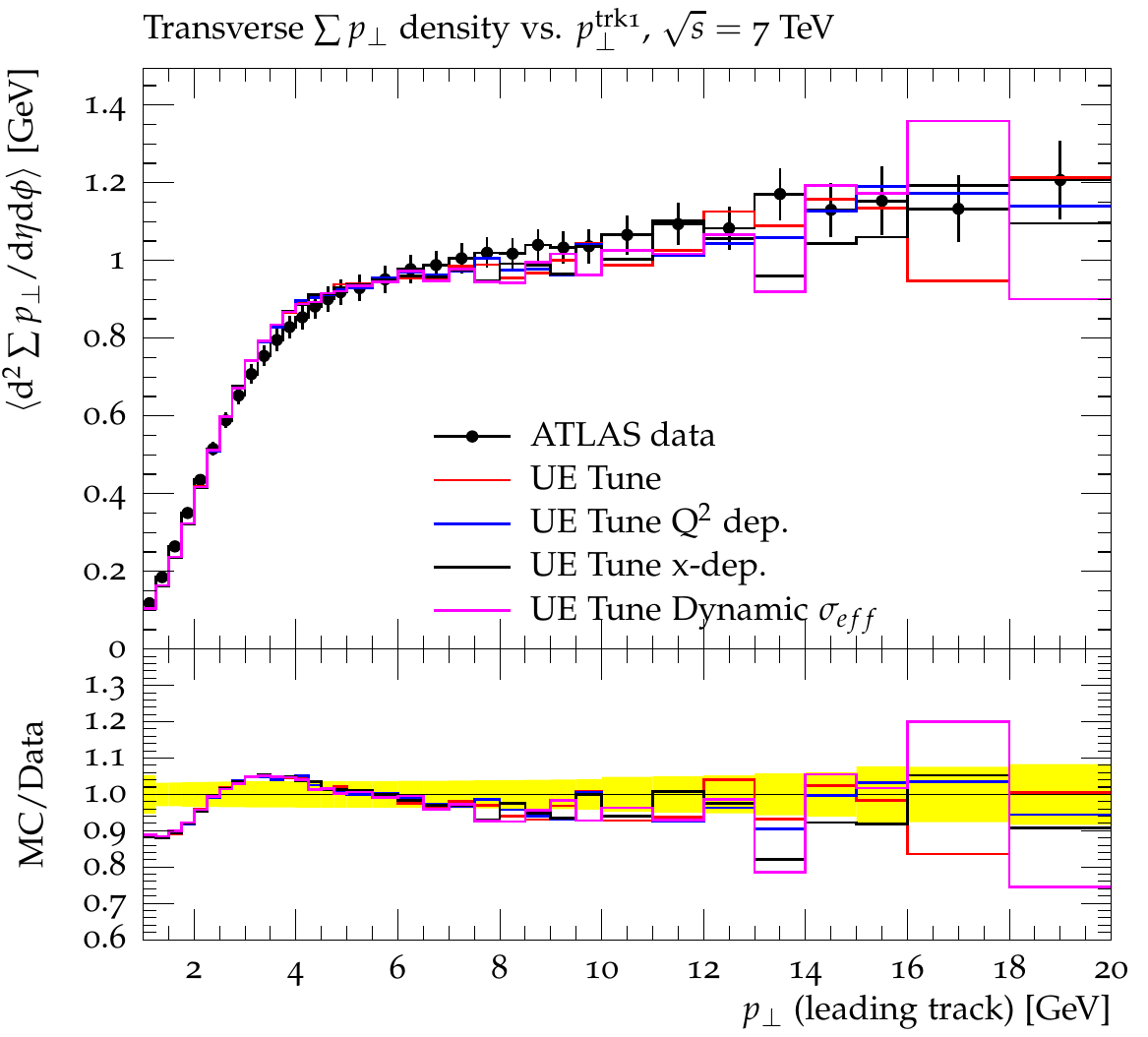}\\
\caption{Charged particle density (left) and \pt sum density (right) as a function of the leading charged particle in the transverse regions, measured by the ATLAS experiment at 7 TeV~\cite{Aad:2010fh}. The data are compared to various predictions: the UE tune with constant \effs\ value (red curve), the UE tune with \effs\ $x$ dependence applied (blue curve), the UE tune with \effs\ scale dependence with $Q_0^2$=1.0 GeV$^2$ applied (black curve) and the UE tune with both \effs\ $x$ and scale dependence with $Q_0^2$=1.0 GeV$^2$ applied (pink curve). The lower panel shows the ratio between the various prediction and the experimental points.}
\label{fig1}
\end{center}
\end{figure}

In this Section, comparisons between UE- and DPS-sensitive measurements at 7 TeV and various predictions are shown. Figure \ref{fig1} shows comparisons to ATLAS data~\cite{Aad:2010fh} on charged particle multiplicity and \pt\ sum in the transverse region as a function of the leading charged particle \pt. Note that these are the observables which have been used in the fitting procedure for the determination of the ``UE tune''.
The measurement is well reproduced by all considered predictions with discrepancies of only up to 10\% in the high-\pt\ region (\pt\ $>$ 10 GeV). The intermediate \pt\ region (2 $<$ \pt\ $<$ 10 GeV) is very well reproduced, while all predictions underestimate the first bins at \pt\ $>$ 2 GeV. This effect might be due to a not optimal simulation of diffraction in \textsc{Pythia}~8. However, no relevant differences are observed for the different \effs\ models.

\begin{figure}[htbp]
\begin{center}
\includegraphics[scale=0.65]{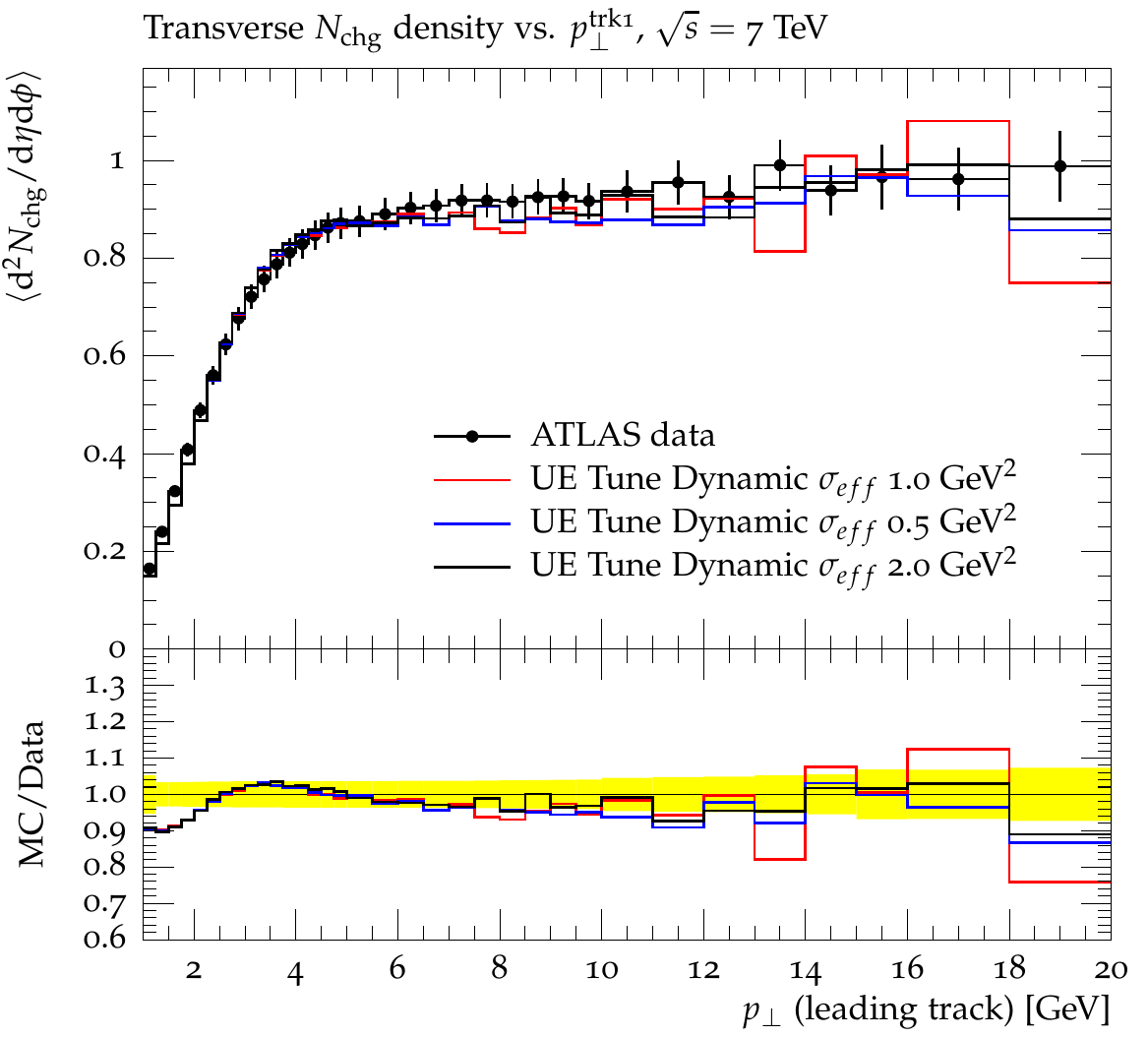}
\includegraphics[scale=0.65]{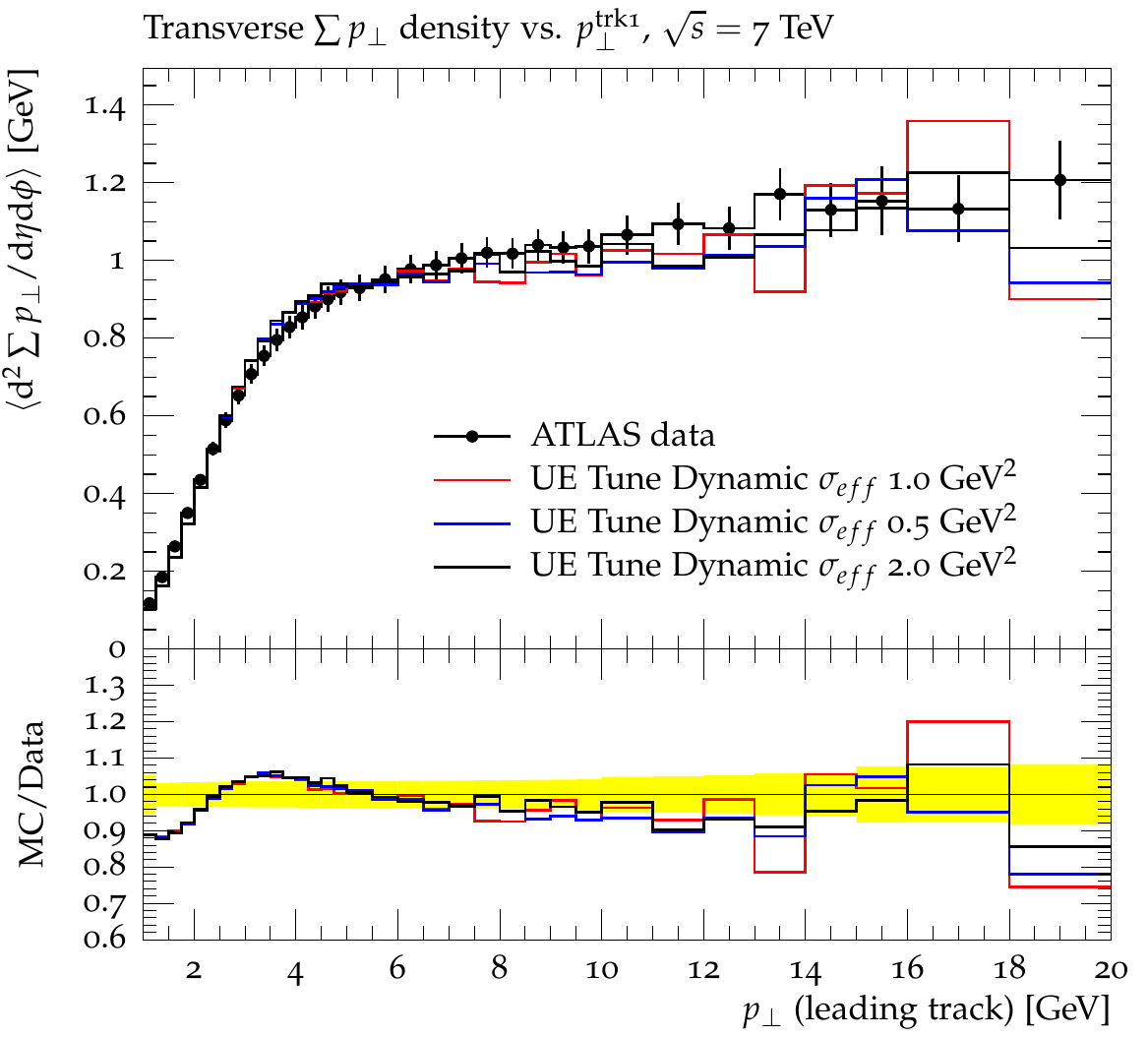}\\
\caption{Charged particle density (left) and \pt sum density (right) as a function of the leading charged particle in the transverse regions, measured by the ATLAS experiment at 7 TeV~\cite{Aad:2010fh}. The data are compared to various predictions obtained with the ``UE tune'' where both the x and scale dependence have been applied for \effs\ with $Q_0^2$ equal to 1.0 (red curve), 0.5 (blue curve) and 2.0 (black curve) GeV$^2$. The lower panel shows the ratio between the various prediction and the experimental points.}
\label{fig2}
\end{center}
\end{figure}

In Figure \ref{fig2}, predictions obtained with different values of the scale $Q_0^2$ values are shown. All predictions are able to reproduce the measurement at the same good level. From this study, one may conclude that the UE data are not sensitive to the different settings of dynamical dependence applied to \effs.

\begin{figure}[htbp]
\begin{center}
\includegraphics[scale=0.65]{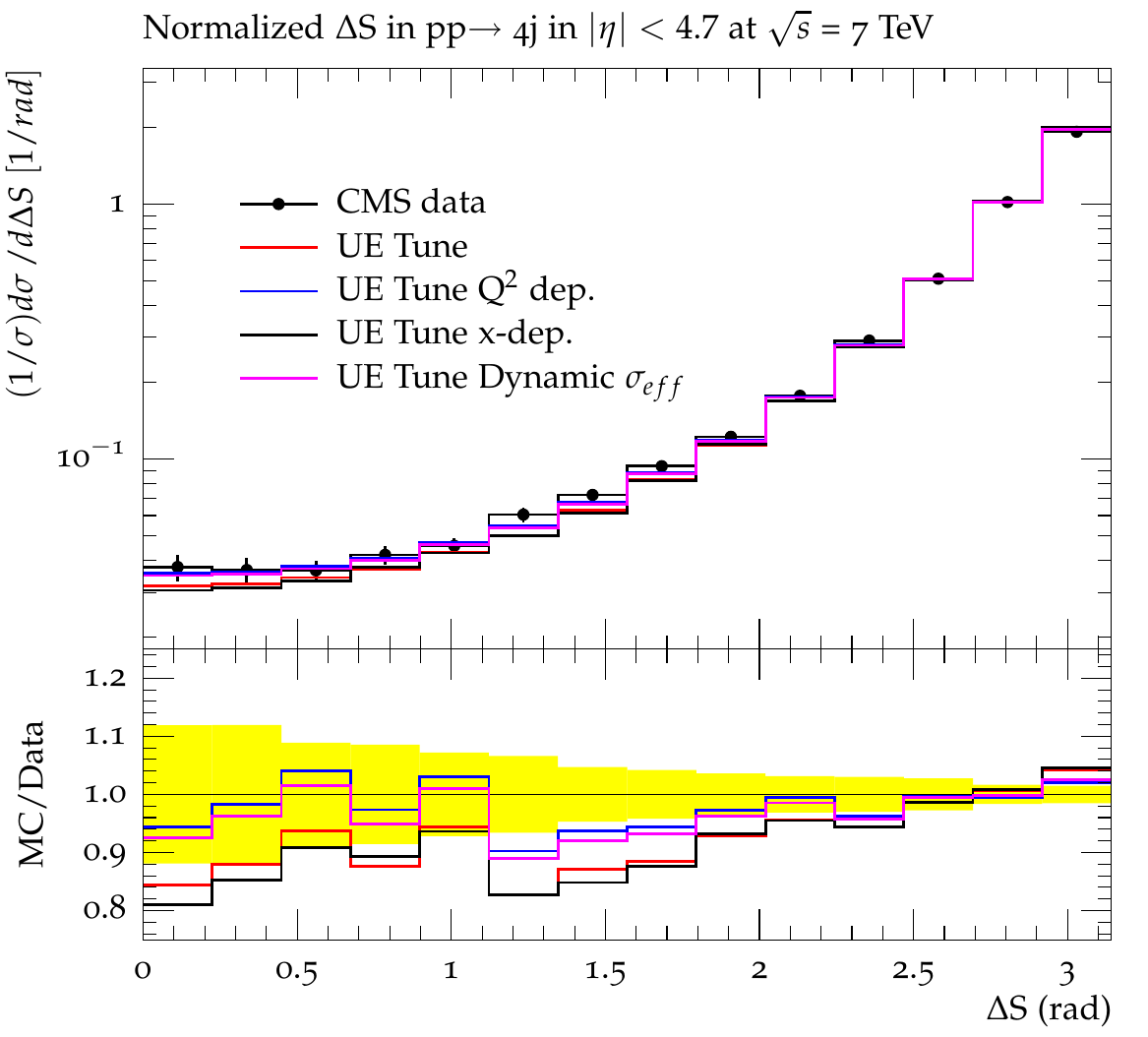}
\includegraphics[scale=0.65]{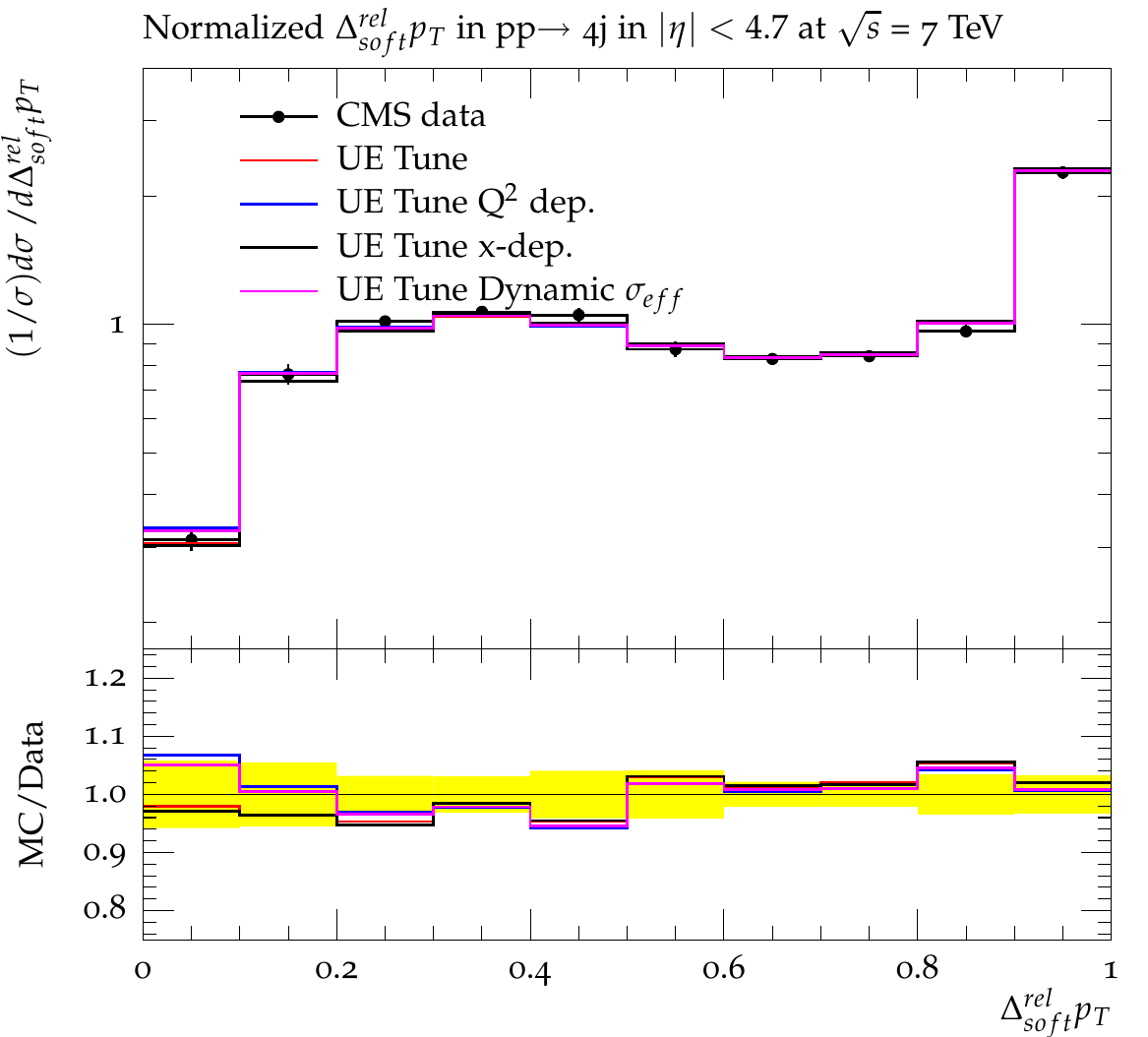}
\caption{Normalized cross section distributions as a function of the correlation observables $\Delta$S (left) and \relpt\ (right) measured in a four-jet scenario by the CMS experiment at 7 TeV~\cite{Chatrchyan:2013qza}. The data are compared to various predictions: the new UE tune (red curve), the new UE tune with the $x$ dependence applied (blue curve), the new UE tune with only the scale dependence with $Q_0^2$=1.0 GeV$^2$ applied (black curve) and the new UE tune with both $x$ and scale dependence with $Q_0^2$=1.0 GeV$^2$ applied (pink curve). The lower panel shows the ratio between the various prediction and the experimental points.}
\label{fig3}
\end{center}
\end{figure}

Figure \ref{fig3} shows predictions with the various \effs\ settings considered previously, compared to the normalized cross section distributions as a function of the correlation observables, $\Delta$S and \relpt, measured in four-jet scenarios~\cite{Chatrchyan:2013qza}. For these variables, the considered models show relevant differences. The static \effs\ dependence (``UE tune'') is not able to properly describe the distribution as a function of $\Delta$S; in particular, the region at low values ($\Delta$S $<$ 2.5), where a DPS contribution is expected, is underestimated by about 10--18\%. By introducing the $x$ dependence for \effs\ (``UE tune x-dep''), the agreement at low values of $\Delta$S does not significantly improve. When the scale dependence of \effs\  is introduced (``UE tune Q$^2$-dep''), the description of the normalized cross section as a function of $\Delta$S gets better with differences not larger than 10\%. The best agreement with the measurement is obtained for predictions where both the $x$ and the Q$^2$ dependence (``UE tune Dynamic \effs'') is included. The normalized cross section as a function of \relpt\ is very well reproduced by all considered predictions. However, it has been already observed in \cite{Chatrchyan:2013qza} that \relpt\ is less sensitive to a DPS contribution than $\Delta$S, which uses information from both jet pairs.

\begin{figure}[htbp]
\begin{center}
\includegraphics[scale=0.65]{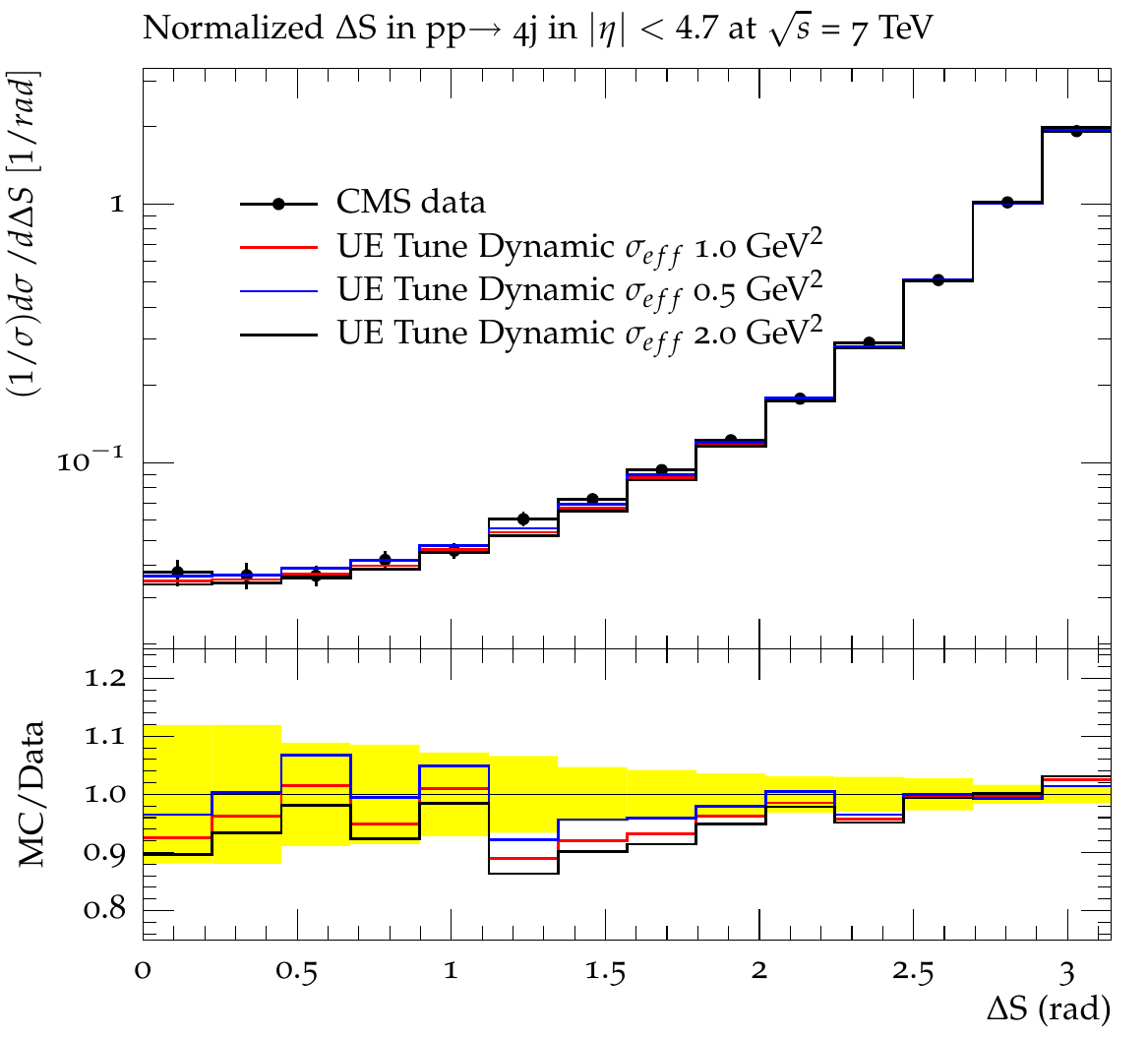}
\includegraphics[scale=0.65]{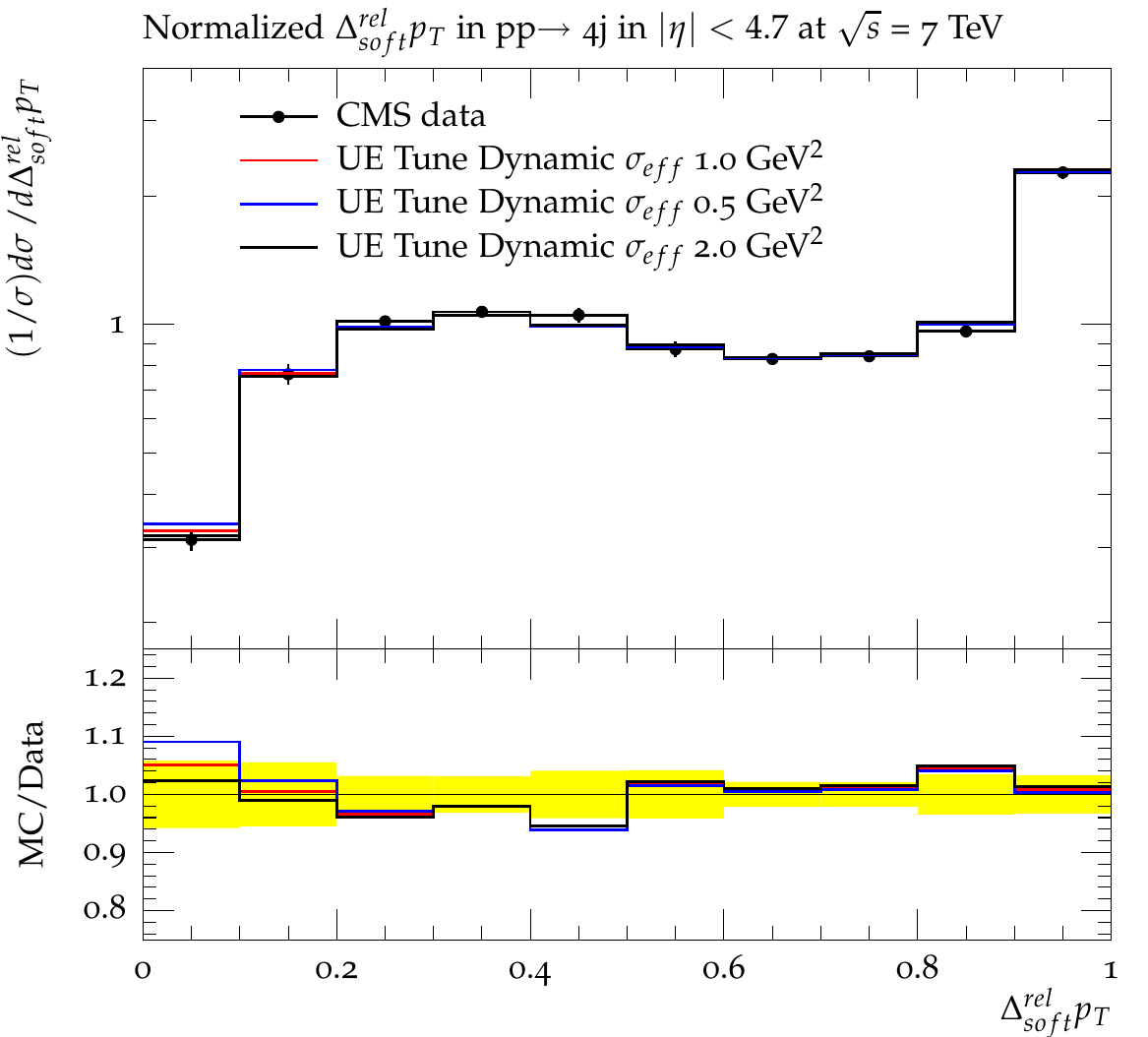}
\caption{Normalized cross section distributions as a function of the correlation observables $\Delta$S (left) and \relpt\ (right) measured in a four-jet scenario by the CMS experiment at 7 TeV~\cite{Chatrchyan:2013qza}. The data are compared to various predictions obtained with the new UE tune where both $x$ and scale dependence have been applied with $Q_0^2$ equal to 1.0 (red curve), 0.5 (blue curve) and 2.0 (black curve) GeV$^2$. The lower panel shows the ratio between the various prediction and the experimental points.}
\label{fig4}
\end{center}
\end{figure}

In Figure \ref{fig4}, predictions obtained with three different values of $Q_0^2$ (0.5, 1.0 and 2.0 GeV$^2$) are compared to the normalized cross section distributions as a function of $\Delta$S and \relpt. A considerable level of agreement for the different settings is obtained. Predictions obtained with $Q_0^2$~=~0.5 GeV$^2$ are in good agreement with the $\Delta$S measurement but overestimate the first bin of \relpt. For $Q_0^2$ = 1 GeV$^2$ and 2 GeV$^2$ the agreement tends to improve for \relpt\ but is worse for $\Delta$S. However, the measurement of the four-jet correlation observables is not able to discriminate the best choice for the value of $Q_0^2$.

\begin{figure}[htbp]
\begin{center}
\includegraphics[scale=0.65]{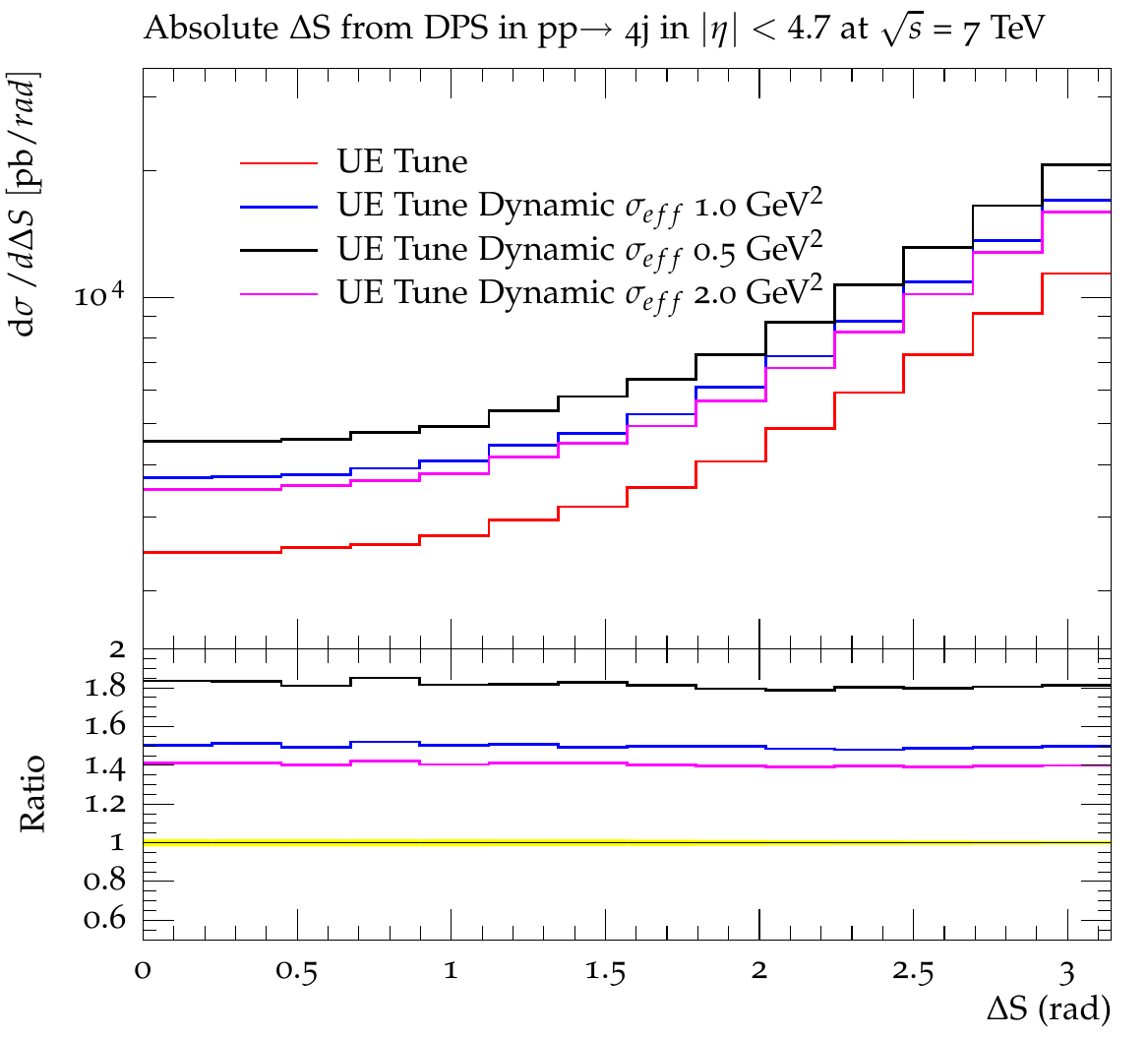}
\includegraphics[scale=0.65]{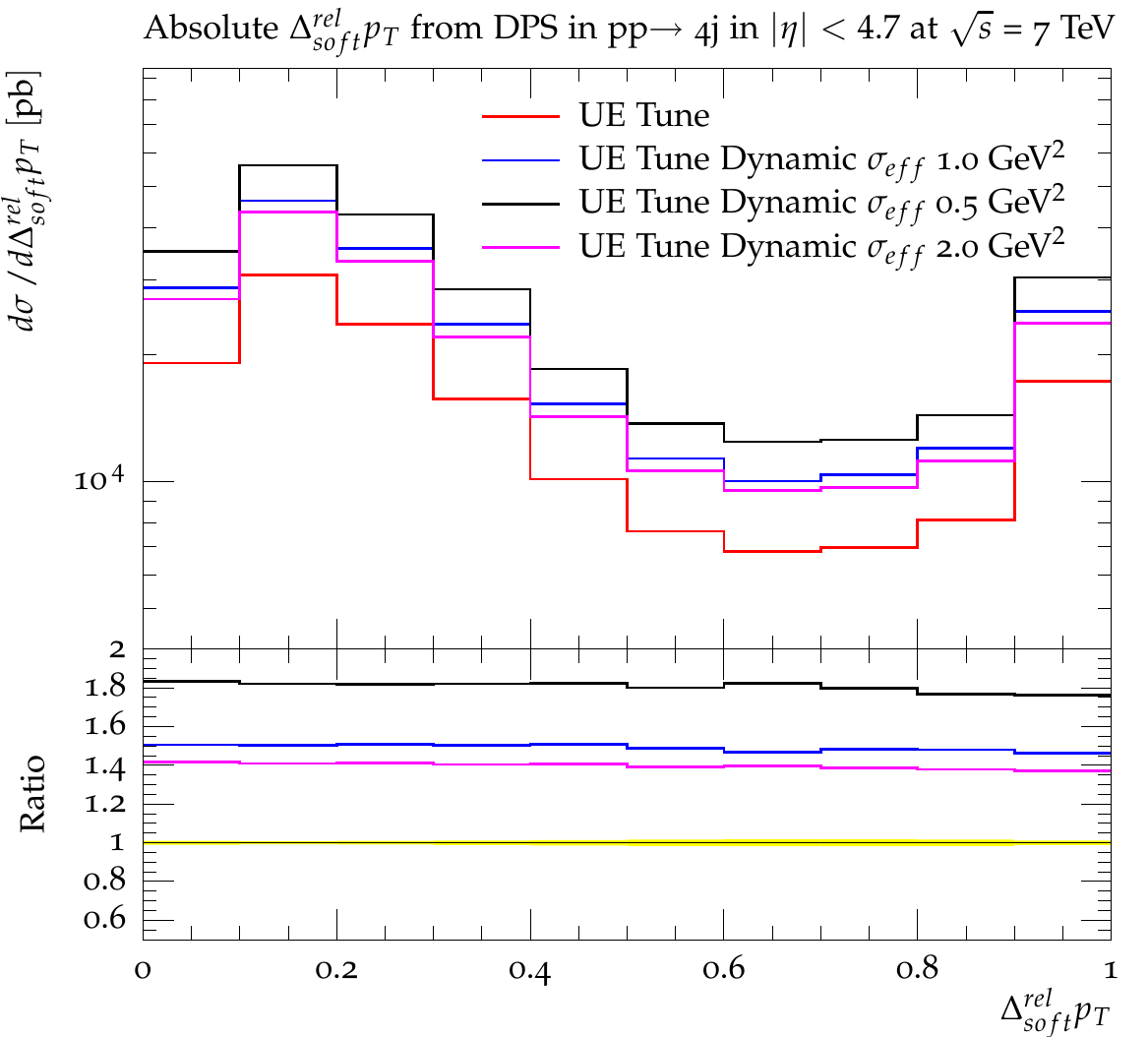}
\caption{Absolute cross section distributions as a function of the correlation observables $\Delta$S (left) and \relpt\ (right), produced via double parton scattering in a four-jet scenario at 7 TeV. Various predictions are shown in the figures: the new UE tune (red curve), the new UE tune with the $x$ dependence applied (blue curve), the new UE tune with only the scale dependence with $Q_0^2$=1.0 GeV$^2$ applied (black curve) and the new UE tune with both $x$ and scale dependence with $Q_0^2$=1.0 GeV$^2$ applied (pink curve). The lower panel shows the ratio between the various predictions and the predictions obtained with the new UE tune.}
\label{fig5}
\end{center}
\end{figure}

In order to isolate the DPS contribution from the background produced by 2$\rightarrow$4 processes, a dedicated event simulation has been performed with \textsc{Pythia}~8. Events with two hard scatterings within the same $pp$ collision are simulated: the first hard scattering is generated with an exchanged transverse momentum between the outgoing partons, $\hat{p}_T$, larger than 45 GeV while for the second one, $\hat{p}_T$ is required to be greater than 15 GeV. Figure \ref{fig5} shows the absolute cross sections predicted by the different settings implemented in the \textsc{Pythia}~8 simulation.

The red curve shows the predictions for the UE tune with a static value of \effs\, while the blue, black and pink lines represent the predictions obtained when implementing the dynamical $x$ and Q$^2$ dependence with $y$ equal to 0.5, 1.0 and 2.0 GeV$^2$, respectively. The highest DPS contribution is observed for $Q_0^2$ = 0.5 GeV$^2$ and it decreases for increasing $Q_0^2$ values. The lowest contribution is observed for the static UE tune when no $x$ and Q$^2$ dependence is applied. The difference between the static and the dynamical \effs\ tune with $Q_0^2$~=~0.5 GeV$^2$ is around 80\%. The different DPS contributions observed among the considered predictions reflect the decreasing \effs\ values for decreasing $Q_0^2$ as a function of the scale of the two scatterings (see Appendix of this paper). No significant differences in the shape of these distributions as a function of $Q_0^2$ are obtained.\\
We observed that predictions of a dynamical \effs\ tune including a $x$ and scale dependence of the transverse parton distribution are fully consistent with experimental data sensitive to moderate and hard MPI. The good agreement obtained for hard MPI is achieved due to contribution of \12 mechanism. The contribution from this mechanism is essentially model independent, except for $Q_0^2$ \cite{BDFS4}, which is the only new fit parameter, which is expected to lie in the $0.5 - 2$ GeV$^2$ range.

\section{Predictions for 14 TeV}
\begin{figure}[htbp]
\begin{center}
\includegraphics[scale=0.65]{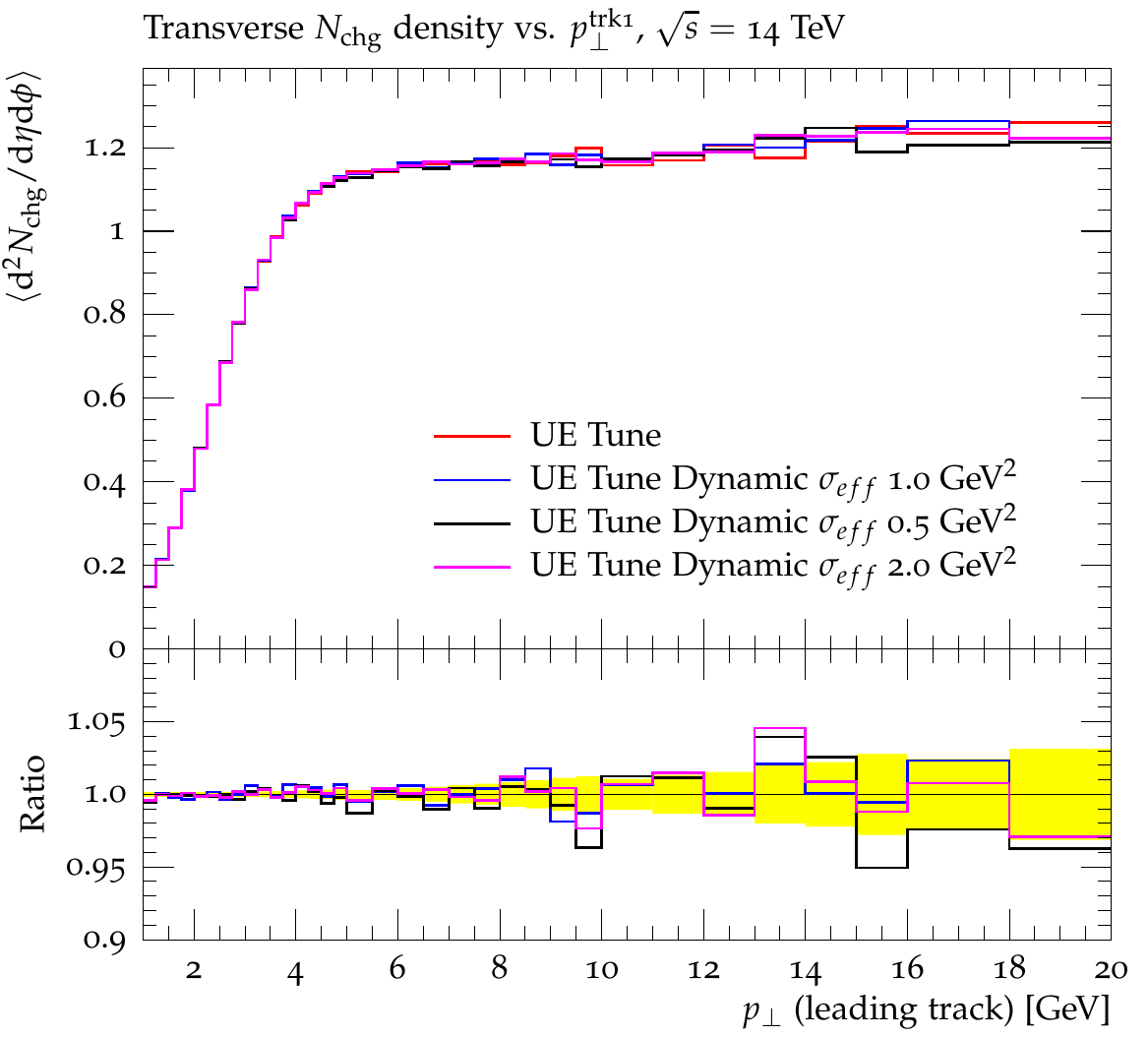}
\includegraphics[scale=0.65]{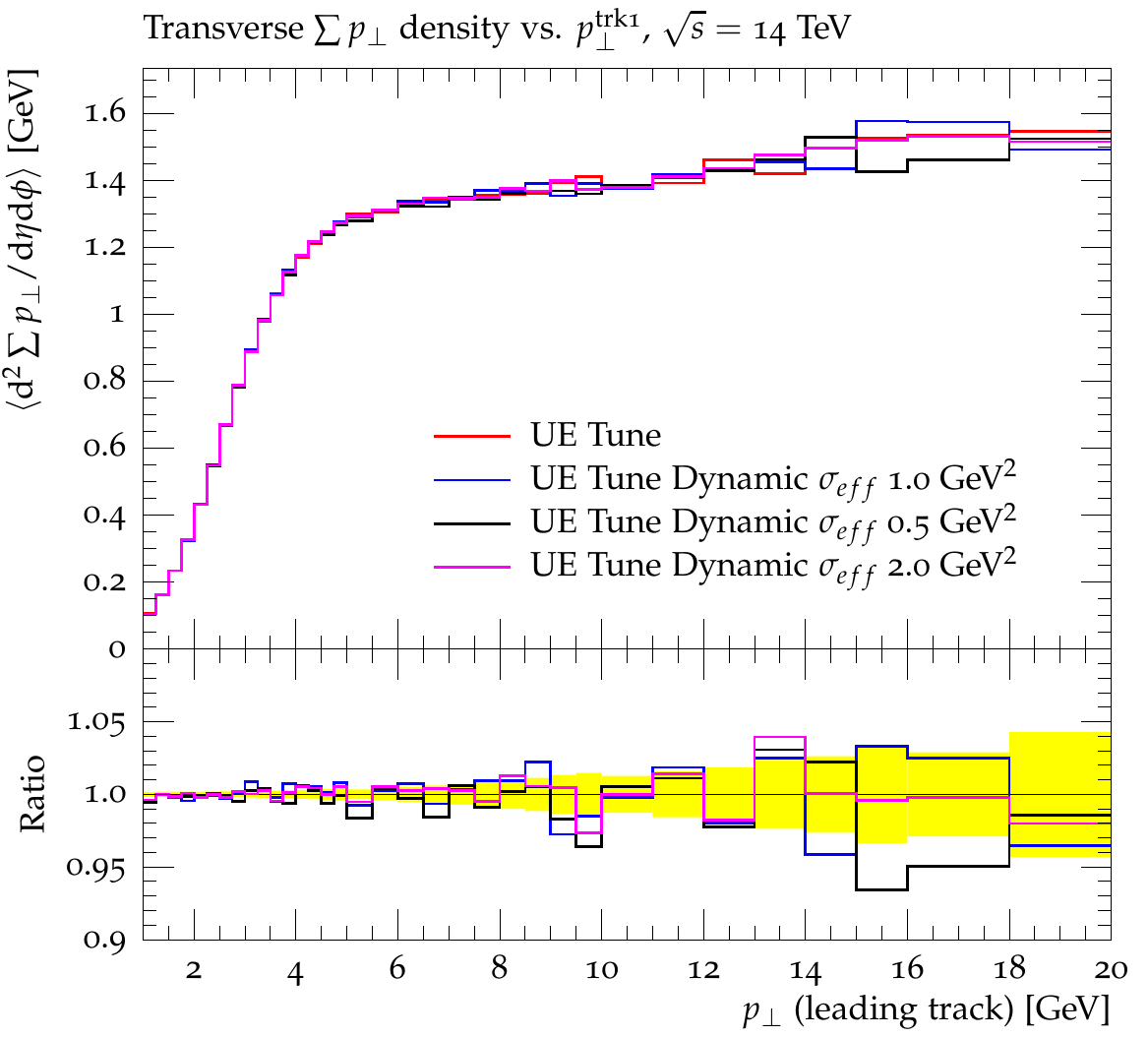}\\
\caption{Charged particle density (left) and \pt sum density (right) as a function of the leading charged particle in the transverse regions at 14 TeV. Various predictions are shown in the figures: the new UE tune (red curve), the new UE tune with both x- and scale-dependence with $Q_0^2$=1.0 GeV$^2$ (blue curve), $Q_0^2$=0.5 GeV$^2$ (black curve) and $Q_0^2$=2.0 GeV$^2$ applied (pink curve). The lower panel shows the ratio between the various prediction and the experimental points.}
\label{fig6}
 \end{center}
\end{figure}

\begin{figure}[htbp]
\begin{center}
\includegraphics[scale=0.65]{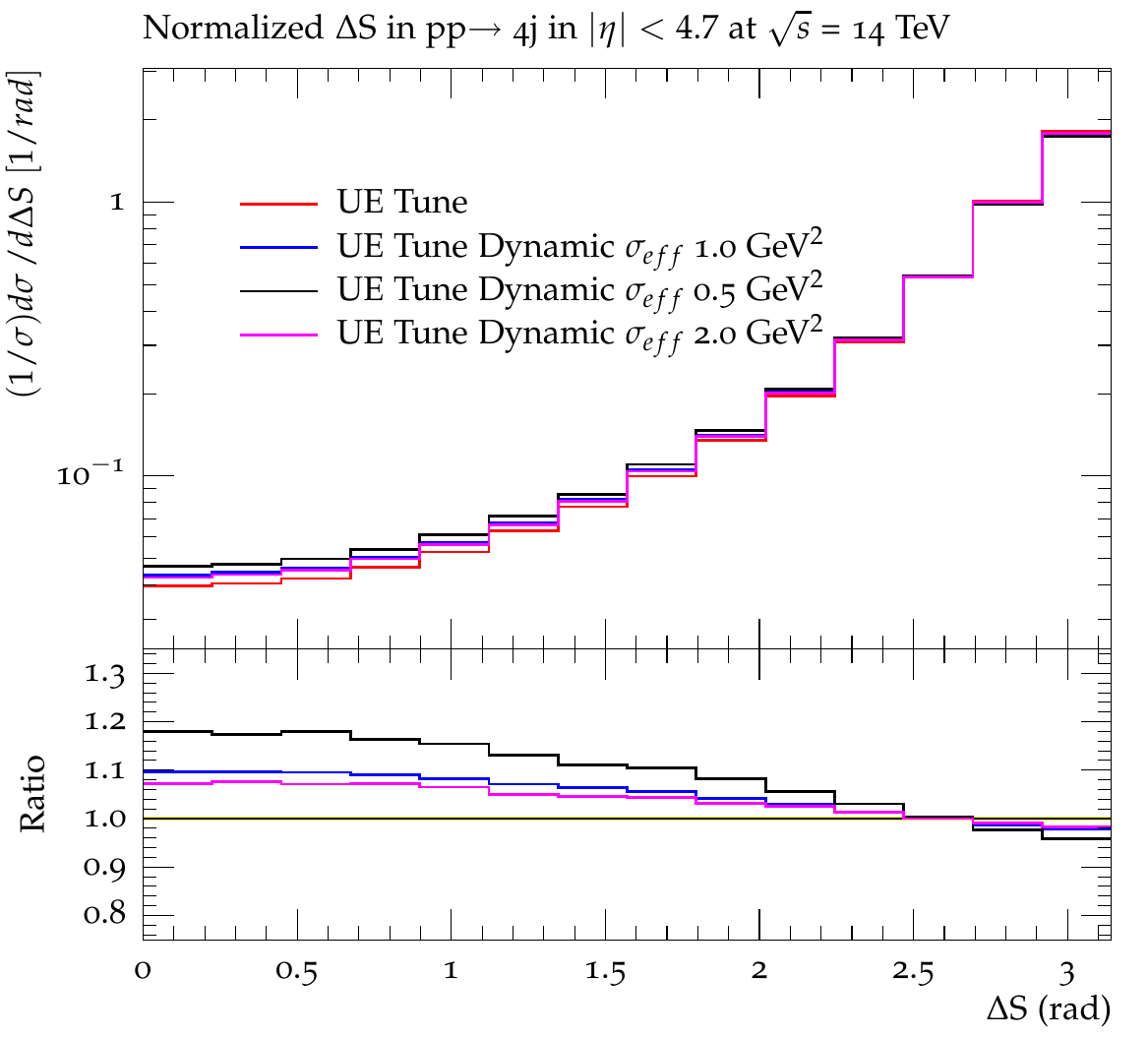}
\includegraphics[scale=0.65]{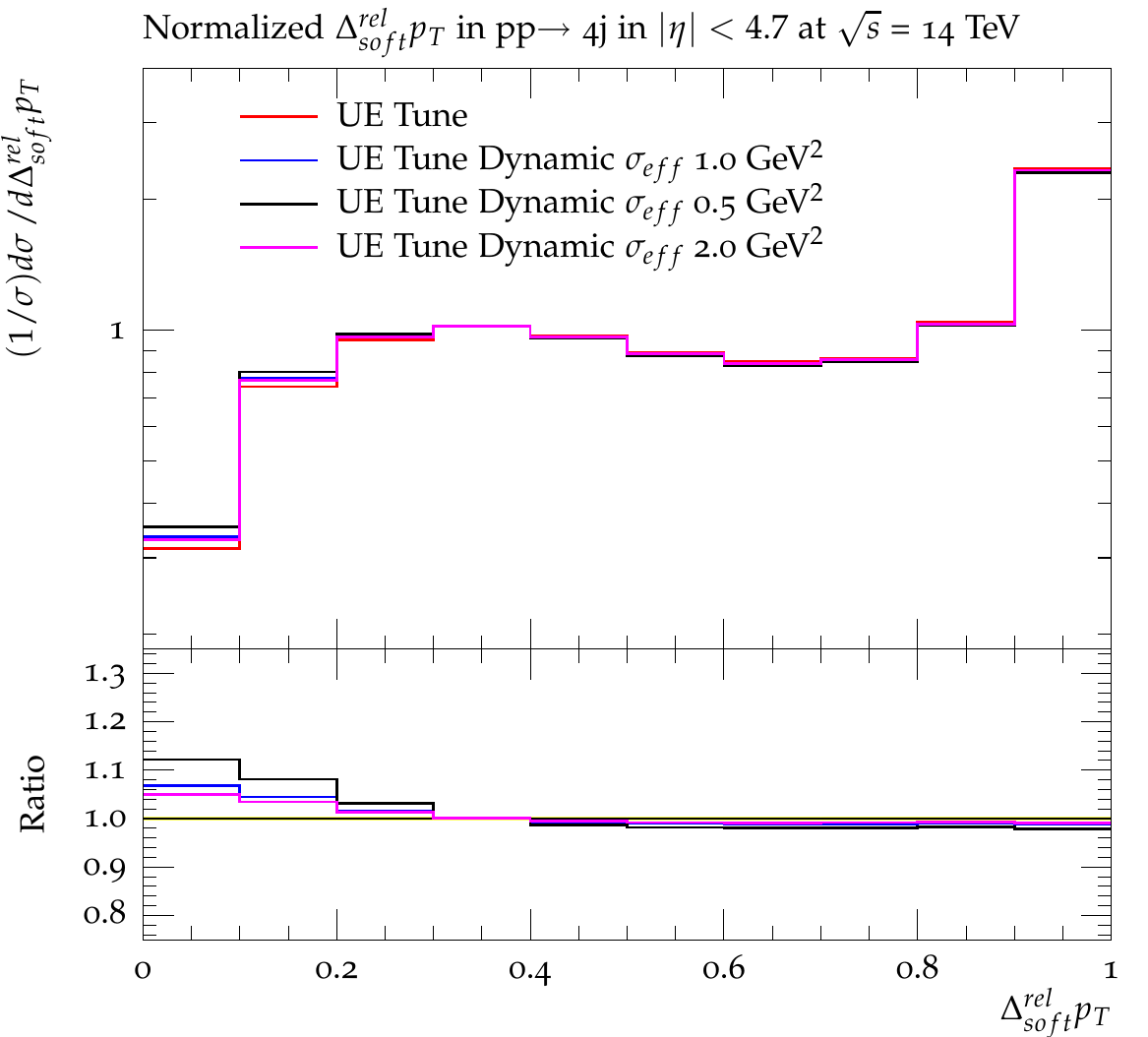}
\caption{Normalized cross section distributions as a function of the correlation observables $\Delta$S (left) and \relpt\ (right) in a four-jet scenario at 14 TeV. Various predictions are shown in the figures: the new UE tune (red curve), the new UE tune with both $x$ and scale dependence with $Q_0^2$=1.0 GeV$^2$ (blue curve), $Q_0^2$=0.5 GeV$^2$ (black curve) and $Q_0^2$=2.0 GeV$^2$ applied (pink curve). The lower panel shows the ratio between the various predictions and the predictions obtained with the new UE tune.}
\label{fig7}
\end{center}
\end{figure}

The dynamical \effs\ dependence has been tested for predictions of UE and DPS observables at a center-of-mass energy of 14 TeV. The $x$ and scale dependence of \effs\ follows respectively Equations \ref{mura} and \ref{scaledep}, similarly as for 7 TeV. Note that the function $R$ in Equation \ref{scaledep} also depends on the center-of-mass energy $\sqrt{s}$ \cite{BDFS2}. Figure \ref{fig6} shows predictions of charged particle density and the \pt\ sum as a function of the leading charged particle \pt, while in Figure \ref{fig7} the normalized cross sections as a function of the four-jet correlation observables, $\Delta$S and \relpt, are presented. The predictions have been obtained by using the UE tune with a static \effs\ value and with a dynamical $x$- and Q$^2$-dependent \effs\ value, with various values for $Q_0^2$: 0.5, 1.0 and 2.0 GeV$^2$.

For each plot, the ratio to the predictions obtained with the UE tune with a constant \effs\ value is shown. While for the considered UE observables, a very small change is observed for the various predictions, larger differences are observed when the four-jet correlation observables are investigated. In particular, tunes with a dynamical \effs\ dependence tend to predict a higher contribution at low $\Delta$S and \relpt\ values. These are the regions where a contribution from DPS is expected. The difference between static and dynamical \effs\ dependence is of up to 15\% for $\Delta$S~$<$~2.0. Predictions with Q$_0^2$ equal to 1.0 and 2.0 GeV$^2$ are very similar to each other, while results obtained with Q$_0^2= 0.5$ GeV$^2$ show a higher contribution at low values of $\Delta$S and \relpt, where the contribution of hard MPI is expected to be relevant.

\section{Comparison with recent Herwig tunes}
\par The calculations described so far in this paper are based on the MPI approach implemented in \textsc{Pythia}. A different approach for the description of MPI is implemented in the \textsc{Herwig}++ event generator~\cite{gieseke1,gieseke2,gieseke3}. Recently, a new tune has been released for the simulation of the UE, labelled as UE-EE-5-CTEQ6L1~\cite{gieseke2}. This tune is very interesting for the purpose of this paper because it is able to simultaneously describe data sensitive to soft MPI and predict a value of $\sigma_{eff}$ of about 15 mb, which is much lower than the one in "Pythia UE tune''. However, the approach of the UE-EE-5-CTEQ6L1 tune is based on a very different picture of both UE and hard MPI than the one discussed in our paper:
\begin{itemize}
\item in \cite{gieseke1,gieseke2,gieseke3}, the mean field approximation is used to describe hard MPI, with parameters related to the transverse parton density distribution obtained through a fit to  hard MPI data. The parametrization of the transverse parton distribution corresponds to a dipole form of the two gluon form factor (Eq.~\ref{d1}) equal to:
\beq
F_{2g}=\left(\frac{1}{1+\Delta^2/m^2_g}\right)^2.
\eeq
The parameter $\mu^2$ \cite{gieseke1,gieseke2,gieseke3} has the same physical interpretation as the parameter $m^2_g$ introduced in \cite{Frankfurt1,BDFS1}, measuring the gluonic radius of the proton. In ``UE Tune Dynamic $\sigma_{eff}$'' developed in this paper, the transverse parton distributions have been determined from HERA data \cite{Frankfurt,Frankfurt1,BDFS1}, having thus the parameter $m^2_g$ as a model-independent input. Comparing $\mu^2$ and $m^2_g$, i.e. comparing the values of the gluonic radii used by tunes UE-EE-5-CTEQ6L1 and ``UE Tune Dynamic $\sigma_{eff}$'', respectively, one gets $\mu^2$ $\sim$ 2$m^2_g$. This means that in the UE-EE-5-CTEQ6L1 tune the gluonic radius of the proton in hard MPI is $\sqrt{2}$ times smaller than the one observed in HERA. In our approach the gluonic radius of the proton is compatible with the one observed at HERA, but in addition to the mean field approximation, a \12 mechanism is included. The contribution of the \12 mechanism to hard MPI is of the same order as of the mean field approximation;
\item in order to describe UE data and to predict $\sigma_{eff}$ around 15 mb, the UE-EE-5-CTEQ6L1 tune uses a color reconnecgtion model developed in \cite{gieseke3}. In such approach one gets the value of $\sim$3.9 GeV for the regularization threshold, $p^0_{T}$, of the partonic cross section. 
    For ``UE Tune Dynamic $\sigma_{eff}$'', the description of UE data and the corresponding parameters are similar to ``Pythia UE tune''. In particular, the value of $p^0_{T}$ implemented in ``UE Tune Dynamic $\sigma_{eff}$'' is $\sim$2.68 GeV (see Table \ref{table1});
\item the MPI model implemented in \textsc{Herwig}++ does not lead to any transverse dependence for the value of $\sigma_{eff}$, which is taken as a constant as a function of the scale of the secondary hard scattering,
    in difference to the current approach.
\end{itemize}

\begin{figure}[htbp]
\begin{center}
\includegraphics[scale=0.65]{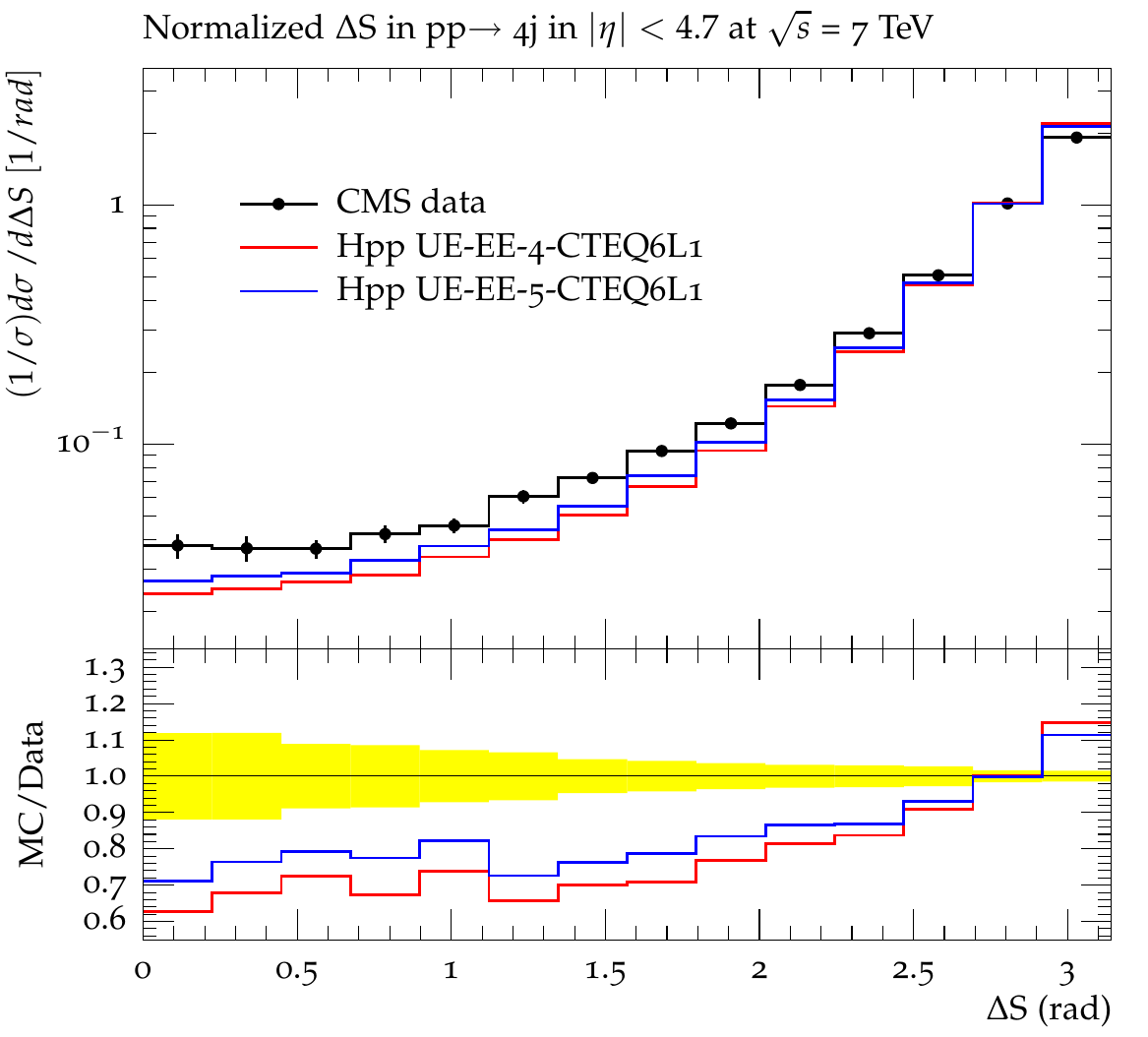}
\includegraphics[scale=0.65]{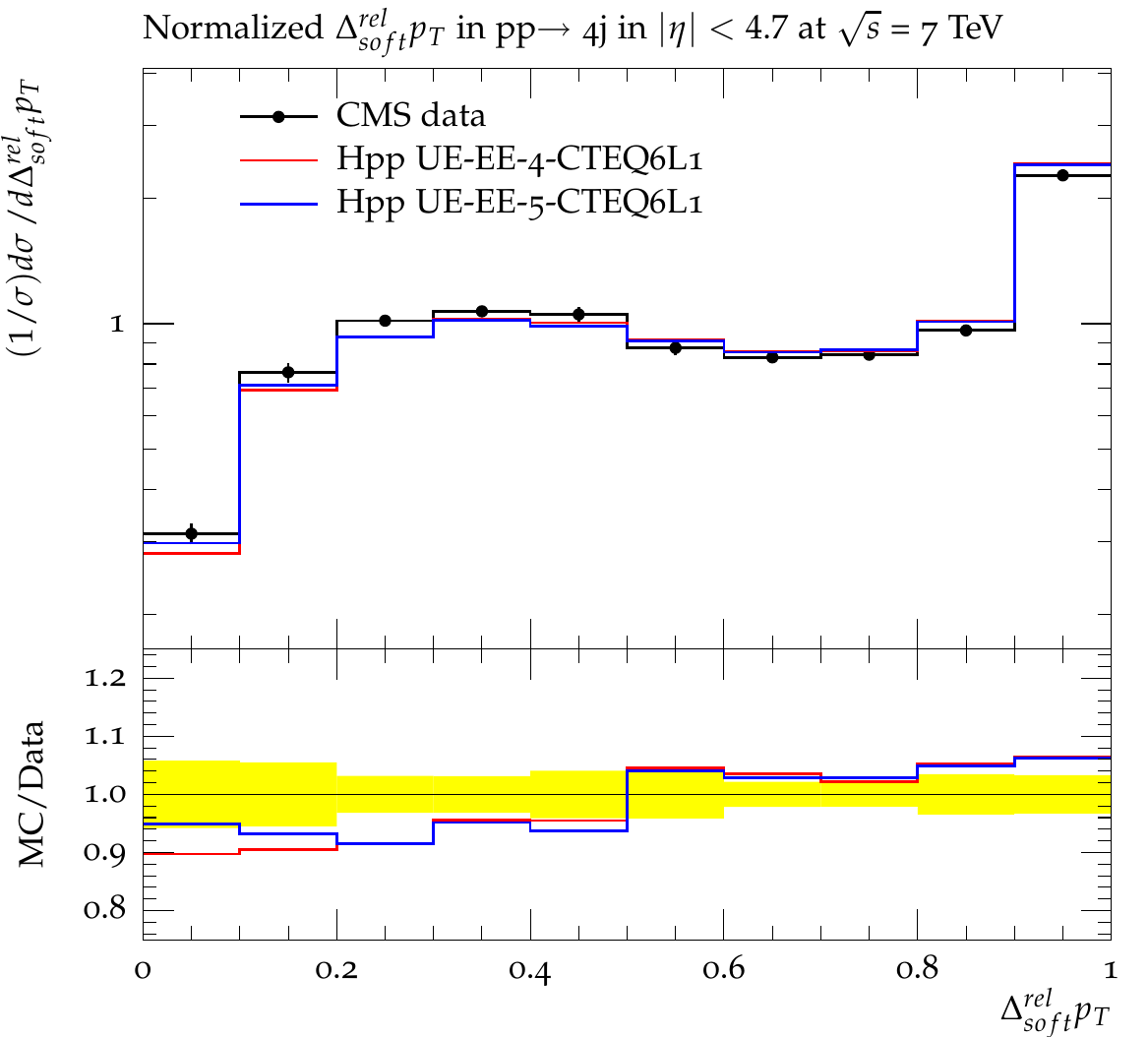}
\caption{Normalized cross section distributions as a function of the correlation observables $\Delta$S (left) and \relpt\ (right) measured in a four-jet scenario by the CMS experiment at 7 TeV~\cite{Chatrchyan:2013qza}. The data are compared to predictions obtained with \textsc{Herwig}~++ tune UE-EE-4-CTEQ6L1 and tune UE-EE-5-CTEQ6L1. The lower panel shows the ratio between the various prediction and the experimental points.}
\label{fig1Herwig}
\end{center}
\end{figure}

Predictions of the two described \textsc{Herwig}++ tunes have been compared to data sensitive to hard MPI. Figure \ref{fig1Herwig} shows predictions of the old UE-EE-4-CTEQ6L1~\cite{gieseke2} and UE-EE-5-CTEQ6L1 tunes, compared to the normalized distributions as a function of the correlation observables, $\Delta$S and $\Delta^{rel}_{soft}p_T$, measured by CMS in four-jet final states at 7 TeV \cite{Chatrchyan:2013qza}. Predictions from both tunes do not give a good description of the experimental data; UE-EE-5-CTEQ6L1 tune performs better than UE-EE-4-CTEQ6L1 but differences of around 20--30\% with the data are observed for values of $\Delta$S smaller than 2.5.

In conclusion, the approach used by the ``UE tune Dynamic \effs'' developed in this paper and by the \textsc{Herwig}++ UE-EE-5-CTEQ6L1 tune \cite{gieseke2}, are rather different and are based on a different picture of both UE and hard MPI. In ``UE tune Dynamic \effs'', the emerging treatment of UE is quite close to mean field approach based on transverse parton densities determined from HERA, and ladder splittings (\12 mechanisms) become important in the description of  processes with hard MPI. In the approach of \textsc{Herwig}++ UE-EE-5-CTEQ6L1 tune, soft and hard MPI are both described in mean field approach, but with a gluon radius of about 1.4 times smaller than the one obtained from exclusive diffraction measurements at HERA, and a new color reconnection model. No \12 mechanism is included.  We believe, that additional experimental data sensitive to soft and hard MPI will be able in the future to further constrain and eventually discriminate the two approaches.

\section{Conclusions}
\par We have developed a new tune ``UE tune Dynamic \effs''\footnote{The code in \textsc{RIVET} of the two analyses, UE and four-jet measurements, implementing the described event reweighting, can be obtained at the following link: http://desy.de/$\sim$gunnep/SigmaEffectiveDependence/.}. The code
 modifies the treatment of hard Multiple Parton Interactions (MPI) in \textsc{Pythia}~8, leading to an improvement in the description of experimental data. We do not change the Monte Carlo code of \textsc{Pythia}, but we rather use the results of the MPI simulation on an event-to-event basis, so that \12 mechanisms are included.
  The tune uses a fit to Underlying Event (UE) data in order to extract the parameters relative to soft MPI and includes values of \effs, which contain the \12 mechanism. They are calculated directly in the mean field + pQCD approach, as discussed in~\cite{BDFS4}. The dynamical dependence of \effs\ is not derived from a process-dependent fit of the experimental data, but is directly obtained from theoretical calculations~\cite{BDFS1,BDFS2,BDFS3,BDFS4}. For the parameter $Q_0^2$, that separates soft and hard scales, we have considered a range of values $0.5<Q_0^2<2$ GeV$^2$. At present, the accuracy of the experimental data does not allow to carry a more precise determination, although the central values of the measured observables are better described by $0.5<Q_0^2<1$ GeV$^2$. We observe that predictions from such tune are in good agreement with experimental measurements at 7 TeV, and for the first time give a consistent description of MPI at both moderate (UE) and hard scales. The results for UE are close to mean field approximation values, as anticipated in \cite{BDFS3}. The additional transverse scale dependence of \effs, relative to mean field approach, due to \12 mechanism, is essential  for a unified description of UE and hard MPI.

 Predictions, obtained with the new tune for proton-proton collisions at 14 TeV, which are expected to happen within the next LHC phase, are also presented.

\section*{Acknowledgements}
We thank M. Strikman, Y. Dokshitzer, H. Jung and S. Dooling for very useful discussions and reading the manuscript.

\appendix
\section{\effs\ dependence at different energies for various scale and longitudinal momentum fraction choices}
\begin{figure}[htbp]
\begin{center}
\includegraphics[scale=0.45]{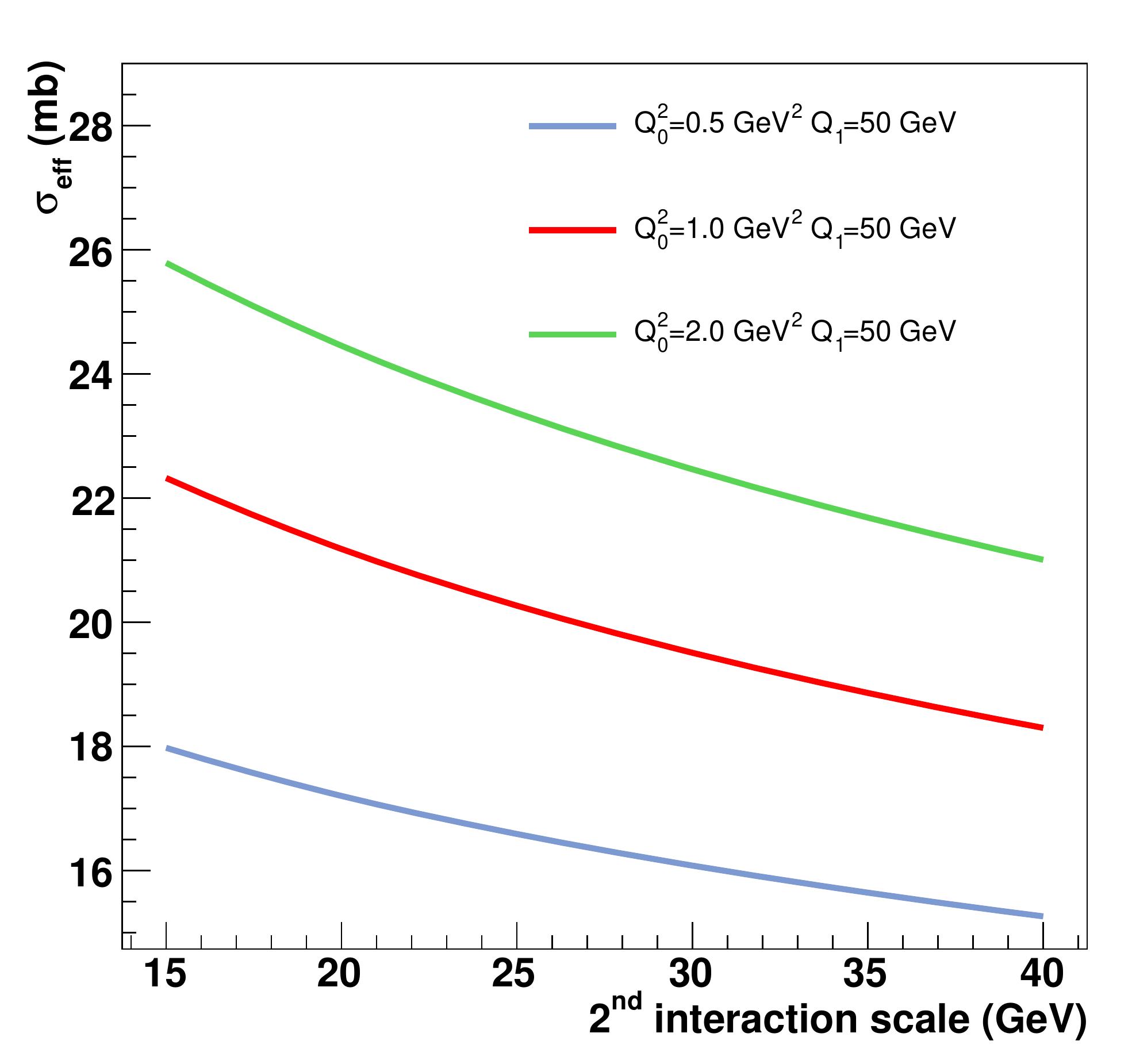}
\caption{Values of \effs\ as a function of the scale of the 2$^{nd}$ interaction for different scales of the first interaction, Q$_1$, and different choices of $Q_0^2$. The values of the longitudinal momentum fractions correspond to the maximal transverse momentum exchange.}
\label{fig1app}
\end{center}
\end{figure}

In this Section, a closer look at the \effs\ dependence on scale, longitudinal momentum fraction and collision energy is provided. Figure \ref{fig1app} shows the values of \effs\ as a function of the scale of the 2$^{nd}$ interaction for a scale of the first interaction equal to 50 GeV and different choices of $Q_0^2$ (0.5, 1.0 and 2.0 GeV$^2$). In this study, the longitudinal momentum fractions of the first interaction system has been set to 0.014, corresponding to the maximal transversality regime. The $x$ value relative to the second hard scattering has been also fixed to the maximal transverse momentum exchange. One can see that \effs\ spans over a range of values between 16 and 30 mb, depending on the choice of $Q_0^2$. The value of \effs\ decreases as a function of the scale of the 2$^{nd}$ hard interaction, Q$_2$, showing a difference of about a factor of 1.1-1.2 between Q$_2$ = 15 GeV and Q$_2$ = 40 GeV. A significant dependence of \effs\ on the choice of $Q_0^2$ is also observed. The smallest \effs\ values are obtained for $Q_0^2$=0.5 GeV$^2$, while they increase of roughly a factor of 1.25 and 1.44, for respectively $Q_0^2$=1.0 and $Q_0^2$=2.0 GeV$^2$.

\begin{figure}[htbp]
\begin{center}
\includegraphics[scale=0.45]{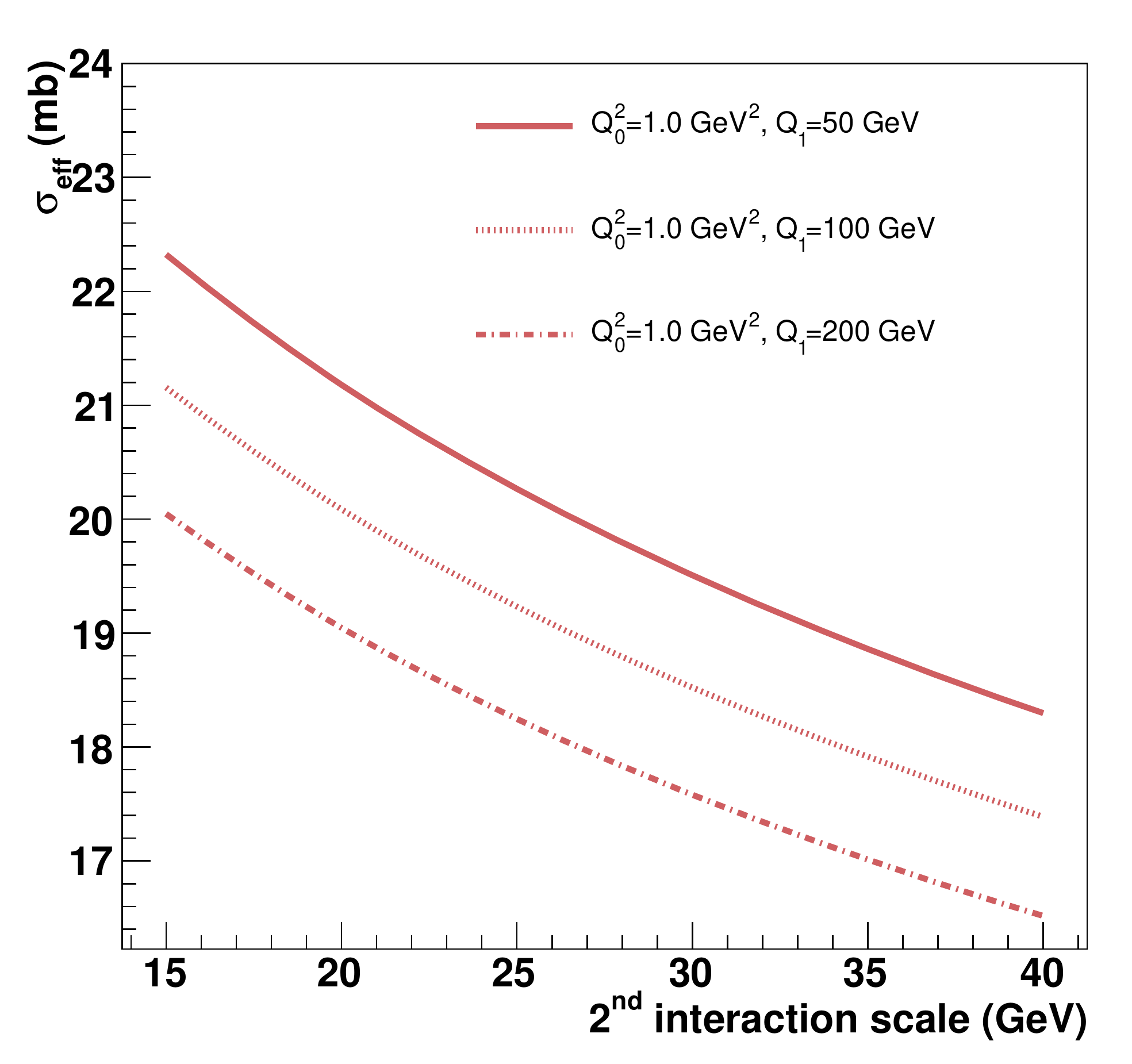}
\caption{Values of \effs\ as a function of the scale of the 2$^{nd}$ interaction for different scales of the first interaction, Q$_1$. The value of $Q_0^2$ has been kept fixed to 1.0 GeV$^2$.}
\label{fig2app}
\end{center}
\end{figure}

\begin{figure}[htbp]
\begin{center}
\includegraphics[scale=0.45]{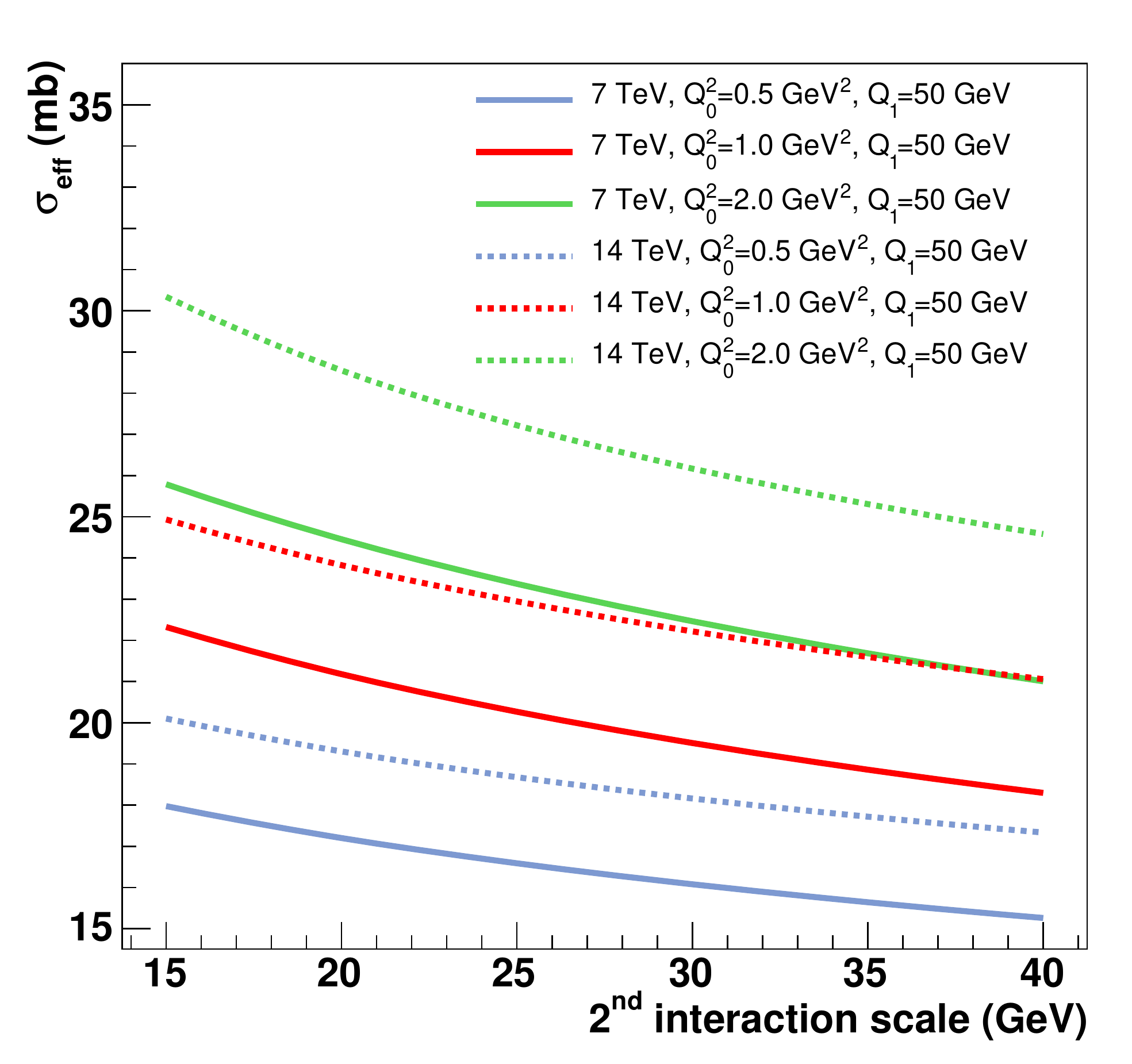}
\caption{Values of \effs\ as a function of the scale of the 2$^{nd}$ interaction at different collision energies at 7 TeV and 14 TeV for first hard interactions occurring at a scale $Q_1$ = 50 GeV. The three values of $Q_0^2$ equal to 0.5, 1.0 and 2.0 GeV$^2$ are considered and the longitudinal momentum fractions of the two dijets correspond to the maximal transverse momentum exchange for both $\sqrt{s}$ = 7 TeV and $\sqrt{s}$ = 14 TeV.}
\label{fig3app}
\end{center}
\end{figure}

In Figure \ref{fig2app}, the \effs\ dependence is studied for various scales of the first interaction (50, 100 and 200 GeV) corresponding to choices of $x_1$ and $x_2$ in the maximum transversality regime, equal to respectively 0.014, 0.028 and 0.056. The values of $x_{3}$ and $x_4$ related to the partons participating in the secondary hard scattering are also set to the maximal exchanged transverse momentum. In this study, only predictions obtained with $Q_0^2$ = 1.0 GeV$^2$ are shown. It is observed that \effs\ does not show a large dependence on the scale of the first interaction: in particular, \effs\ decreases as a function of the scale of the first hard scattering. The three curves are very similar between each other as a function of the scale of the second hard interaction and the difference is less than 1 mb.

Figure \ref{fig3app} considers the \effs\ variation at different collision energies, 7 and 14 TeV, as a function of the scale of the second hard interaction. The three values of $Q_0^2$ equal to 0.5, 1.0 and 2.0 GeV$^2$ are considered. Only scales of the first interaction equal to 50 GeV are examined. The value of \effs\ increases for increasing collision energies. For $Q_0^2$=0.5 and 1.0 GeV$^2$, \effs\ increases of about 2-3 mb for any scale of the second hard scattering, while for $Q_0^2$ = 2.0 GeV$^2$, the increase of \effs\ is larger and it reaches values of up to 4.5 mb at $Q_2$ = 15 GeV.

\bibliography{mpi21}

\begin{thebibliography}{99}
\bibitem{TreleaniPaver82}
  N.\ Paver and D.\ Treleani,
  Nuovo Cim.\  A {\bf 70} (1982) 215.
\bibitem{TreleaniPaver85}
 N.\ Paver and D.\ Treleani,
  Z.\ Phys.\  C {\bf 28}  187 (1985).

\bibitem{mufti} M.\ Mekhfi, Phys. Rev. D{\bf 32}, 2371 (1985).
\bibitem{dDGLAP}
	R. Kirschner,
	Phys.\ Lett.\ {\bf B84}, 266  (1979); \\
      V.P.\ Shelest, A.M.\ Snigirev and G.M.\ Zinovjev,
	Phys.\ Lett.\ {\bf B113},  325  (1982).
	\bibitem{Treleani} A.\ Del Fabbro and D.\ Treleani,
	  Phys.\ Rev.\  D {\bf 61}, 077502 (2000)
	  [arXiv:hep-ph/9911358]; Phys.\ Rev.\  D {\bf 63}, 057901 (2001)
	  [arXiv:hep-ph/0005273];  A.\ Accardi and D.\ Treleani,
	  Phys.\ Rev.\  D {\bf 63}, 116002 (2001)
	  [arXiv:hep-ph/0009234].
	
	\bibitem{Wiedemann} S.\ Domdey, H.J.\ Pirner and U.A.\ Wiedemann,
	  Eur.\ Phys.\ J.\  C {\bf 65}, 153 (2010)
	  [arXiv:0906.4335 [hep-ph]].
	
	
	
	\bibitem{SST}  T.C.\ Rogers, A.M.\ Stasto and M.I.\ Strikman,
	  Phys.\ Rev.\  D {\bf 77}, 114009 (2008)
	  [arXiv:0801.0303 [hep-ph]].
	
	\bibitem{Diehl}
	  M.~Diehl,
	  PoS D {\bf IS2010} (2010) 223
	  [arXiv:1007.5477 [hep-ph]].
		
	\bibitem{DiehlSchafer} M.~Diehl and A.~Schafer,
	  Phys.\ Lett.\  B {\bf 698} (2011) 389
	  [arXiv:1102.3081 [hep-ph]].

	\bibitem{Diehl2} M.\ Diehl, D.\ Ostermeier and A.\ Schafer,
	  JHEP {\bf 1203} (2012) 089
	  [arXiv:1111.0910 [hep-ph]].

	\bibitem{Berger}
	E.L.\ Berger, C.B.\ Jackson, G.\ Shaughnessy,
	  Phys.\ Rev.\  {\bf D81}, 014014 (2010)
	  [arXiv:0911.5348 [hep-ph]].

	\bibitem{Ryskin}  M.G.\ Ryskin and A.M.\ Snigirev,
	  Phys.\ Rev.\ {\bf D83}, 114047 (2011)
	  [arXiv:1103.3495 [hep-ph]].
		
	  \bibitem{stirling} J.R.\ Gaunt and W.J.\ Stirling,
	  JHEP {\bf 1003}, 005 (2010)   [arXiv:0910.4347 [hep-ph]]; \\
	J.R.\ Gaunt, C.H.\ Kom, A.\ Kulesza and W.J.\ Stirling,
	  Eur.\ Phys.\ J.\  C {\bf 69}, 53 (2010)  [arXiv:1003.3953 [hep-ph]].
	
	\bibitem{stirling1} J.R.\ Gaunt and W.J.\ Stirling,
	  JHEP {\bf 1106},  048 (2011) [arXiv:1103.1888 [hep-ph]].
	
	\bibitem{Frankfurt}
	  L.\ Frankfurt, M.\ Strikman and C.\ Weiss,
	  Phys.\ Rev.\  D {\bf 69}, 114010 (2004)
  [arXiv:hep-ph/0311231];
	  Ann.\ Rev.\ Nucl.\ Part.\ Sci.\  {\bf 55}, 403 (2005)
	  [arXiv:hep-ph/0507286].
    \bibitem{Frankfurt1}
  L.~Frankfurt, M.~Strikman and C.~Weiss,
  Phys.\ Rev.\ D {\bf 83} (2011) 054012
  [arXiv:1009.2559 [hep-ph]].
\bibitem{BDFS1}
  B.\ Blok, Yu.\ Dokshitzer, L.\ Frankfurt and M.\ Strikman,
  Phys.\ Rev.\  D {\bf 83}, 071501 (2011)
  [arXiv:1009.2714 [hep-ph]].

  \bibitem{BDFS2} B.\ Blok, Yu.\ Dokshitser, L.\ Frankfurt and M.\ Strikman,
  Eur.\ Phys.\ J.\ C {\bf72}, 1963  (2012)
  [arXiv:1106.5533 [hep-ph]].
\bibitem{BDFS3} B.\ Blok, Yu.\ Dokshitser, L.\ Frankfurt and M.\ Strikman,
 arXiv:1206.5594v1 [hep-ph] (unpublished).
 \bibitem{BDFS4}
 B.~Blok, Y.~Dokshitzer, L.~Frankfurt and M.~Strikman,
  Eur.\ Phys.\ J.\ C {\bf 74} (2014) 2926
  [arXiv:1306.3763 [hep-ph]].
  \bibitem{Gauntnew}
  J.~R.~Gaunt, R.~Maciula and A.~Szczurek,
  arXiv:1407.5821 [hep-ph].
  \bibitem{Gauntadd}
J.R.\ Gaunt,
JHEP {\bf 1301}, 042  (2013)
[arXiv:1207.0480 [hep-ph]]
\bibitem{gieseke1}
 S.~Gieseke, C.~A.~Rohr and A.~Siodmok,
  arXiv:1110.2675 [hep-ph];
  S.~Gieseke, D.~Grellscheid, K.~Hamilton, A.~Papaefstathiou, S.~Platzer, P.~Richardson, C.~A.~Rohr and P.~Ruzicka {\it et al.},
  arXiv:1102.1672 [hep-ph].

\bibitem{gieseke2}
  M.~H.~Seymour and A.~Siodmok,
  JHEP {\bf 1310} (2013) 113
  [arXiv:1307.5015 [hep-ph]].
\bibitem{gieseke3}
  S.~Gieseke, C.~Rohr and A.~Siodmok,
  Eur.\ Phys.\ J.\ C {\bf 72} (2012) 2225
  [arXiv:1206.0041 [hep-ph]].
\bibitem{Sjodmok}
M.\ B\"ahr, M.\ Myska, M.H. Seymour and A.\ Siodmok,
arXiv: 1302.4325 [hep-ph]

\bibitem{Sjostrand:2007gs}
  T.~Sj\"ostrand, S.~Mrenna and P.~Z.~Skands,
  Comput.\ Phys.\ Commun.\  {\bf 178} (2008) 852
  [arXiv:0710.3820 [hep-ph]].

\bibitem{Corke:2011yy}
  R.~Corke and T.~Sj\"ostrand,
  JHEP {\bf 1105} (2011) 009
  [arXiv:1101.5953 [hep-ph]].

\bibitem{Corke:2010yf}
  R.~Corke and T.~Sj\"ostrand,
  JHEP {\bf 1103} (2011) 032
  [arXiv:1011.1759 [hep-ph]].
	

\bibitem{Herwig}
  J.~M.~Butterworth, J.~R.~Forshaw and M.~H.~Seymour,
  Z.\ Phys.\ C {\bf 72} (1996) 637
  [hep-ph/9601371].
	
	\bibitem{Lund}
	  C.\ Flensburg, G.\ Gustafson, L.\ Lonnblad and A.\ Ster,
	  arXiv:1103.4320 [hep-ph].

\bibitem{Tevatron1}F.\ Abe {\it et al.}  [CDF Collaboration],
  Phys.\ Rev.\  D {\bf 56}, 3811 (1997).

\bibitem{Tevatron2}  V.M.\ Abazov {\it et al.}  [D0 Collaboration],
  Phys.\ Rev.\  D {\bf 81}, 052012 (2010).

\bibitem{Tevatron3}
V.M.\ Abazov {\it et al.}  [D0 Collaboration],
  Phys.\ Rev.\  D {\bf 83}, 052008 (2011).
\bibitem{Atlas}
 G.~Aad {\it et al.}  [ATLAS Collaboration],
 New J.\ Phys.\  {\bf 15} (2013) 033038
 [arXiv:1301.6872 [hep-ex]].
 \bibitem{cms1}
  S.~Chatrchyan {\it et al.}  [CMS Collaboration],
  JHEP {\bf 1403} (2014) 032
  [arXiv:1312.5729 [hep-ex]].
\bibitem{cms2}
  S.~Chatrchyan {\it et al.}  [CMS Collaboration],
  JHEP {\bf 1312} (2013) 030
  [arXiv:1310.7291 [hep-ex]].
\bibitem{Chatrchyan:2013qza}
  S.~Chatrchyan {\it et al.}  [CMS Collaboration],
  Phys.\ Rev.\ D {\bf 89} (2014) 9,  092010
  [arXiv:1312.6440 [hep-ex]].
\bibitem{Aad:2010fh}
  G.~Aad {\it et al.}  [ATLAS Collaboration],
  Phys.\ Rev.\ D {\bf 83} (2011) 112001
  [arXiv:1012.0791 [hep-ex]].
  \bibitem{DiehlS} M.~Diehl,
  Phys.\ Rept.\  {\bf 388} (2003) 41
  [hep-ph/0307382].
  \bibitem{radyushkin}A.~V.~Belitsky and A.~V.~Radyushkin,
  Phys.\ Rept.\  {\bf 418} (2005) 1
  [hep-ph/0504030].
  \bibitem{gunnelini} P. Gunnellini, talk at 6th MPI@LHC symposium, Krakow, November 2014.
  \bibitem{Buckley:2010ar}
  A.~Buckley, J.~Butterworth, L.~Lonnblad, D.~Grellscheid, H.~Hoeth, J.~Monk, H.~Schulz and F.~Siegert,
  Comput.\ Phys.\ Commun.\  {\bf 184} (2013) 2803
  [arXiv:1003.0694 [hep-ph]].

\bibitem{Buckley:2009bj}
  A.~Buckley, H.~Hoeth, H.~Lacker, H.~Schulz and J.~E.~von Seggern,
  Eur.\ Phys.\ J.\ C {\bf 65} (2010) 331
  [arXiv:0907.2973 [hep-ph]].

\bibitem{Andersson:1998tv}
  B.~Andersson,
  Camb.\ Monogr.\ Part.\ Phys.\ Nucl.\ Phys.\ Cosmol.\  {\bf 7} (1997) 1.
\bibitem{Jung}A.~Grebenyuk, F.~Hautmann, H.~Jung, P.~Katsas and A.~Knutsson,
  Phys.\ Rev.\ D {\bf 86} (2012) 117501
  [arXiv:1209.6265 [hep-ph]].
\bibitem{strikmantalk} M. Strikman, talk at Kyoto University Colloqium,
Kyoto, February 24, 2014.
\bibitem{Pumplin:2002vw}
  J.~Pumplin, D.~R.~Stump, J.~Huston, H.~L.~Lai, P.~M.~Nadolsky and W.~K.~Tung,
  JHEP {\bf 0207} (2002) 012
  [hep-ph/0201195].

\bibitem{Skands:2014pea}
  P.~Skands, S.~Carrazza and J.~Rojo,
  Eur.\ Phys.\ J.\ C {\bf 74} (2014) 8,  3024
  [arXiv:1404.5630 [hep-ph]].


\end{thebibliography}

\end{document}